\documentclass[twocolumn,showpacs,preprintnumbers,amsmath,amssymb,floatfix,prb]{revtex4}

\usepackage{graphicx}
\usepackage{dcolumn}
\usepackage{bm}
\usepackage{longtable}

\input pstricks
\input pst-node
\input pst-plot
\input pst-eps
\input epsf

\def\cE{{\cal E}}
\def\cD{{\cal D}}

\def\bR{{\mathbb R}}
\def\bZ{{\mathbb Z}}
\def\bS{{\mathbb S}}
\def\tR{\bR^3}

\def\bT{{\mathbb T}}
\def\tT{\bT^3}

\def\bp{\bm p}
\def\bq{\bm q}
\def\bE{\bm E}
\def\bH{\bm H}
\def\bM{\bm L}
\def\bL{\bm L}

\begin{document}

\font\fourPalatino  = pplr  at 4truept
\font\sixPalatino   = pplr  at 6truept
\font\eightPalatino = pplr  at 8truept
\font\tenPalatino   = pplr  at 10truept
\font\tenPalatinoIt = pplri at 10truept

\title{
First-principles generation of Stereographic Maps\\ 
for high-field magnetoresistance in normal metals:\\
an application to Au and Ag.
}
\author{Roberto De Leo}
  \email{roberto.deleo@ca.infn.it}
  \altaffiliation[Also at ]{Physics Department, University of Cagliari, Italy}
\affiliation{%
Istituto Nazionale di Fisica Nucleare, sez. di Cagliari\\
Cittadella Universitaria, I-09042 Monserrato (CA), Italy\\
}
\pacs{72.15.Gd,02.40.Re,03.65.Sq}
\date{\today}%
\begin{abstract}
About thirty high-field magnetoresistance Stereographic Maps have been 
measured for metals between Fifties and Seventies but no way was known
till now to compare these complex experimental data with first-principles
computations. We present here the method we developed to generate 
Stereographic Maps directly from a metal's Fermi Surface, based on the 
Lifshitz model and the recent advances by S.P. Novikov and his pupils.
As an application, we test the method with an interesting toy model 
and then with Au and Ag.
\end{abstract}

\maketitle

\section{Introduction}

The relevance of the geometry and topology of the Fermi Surface (FS) in physical 
phenomena is well known since Thirties, when Justi and Scheffers showed evidences
that the Fermi Surface of Gold is open \cite{JS37}.
\par
One of the most striking examples of such phenomena is the behaviour of 
magnetoresistance in monocrystals at low temperatures in high magnetic fields.
Evidences for this effect was first discovered theoretically by Lifshitz and his 
Karkov school in Fifties \cite{LAK56} by studying galvanometric effects in metals
without any special assumption for the electron energy spectrum, and it was
verified experimentally by Gaidukov {\sl et al.} shortly afterwards \cite{Ga60,AG62c}.
\par
\begin{figure*}
\includegraphics[width=5cm, bb=230 540 382 675]{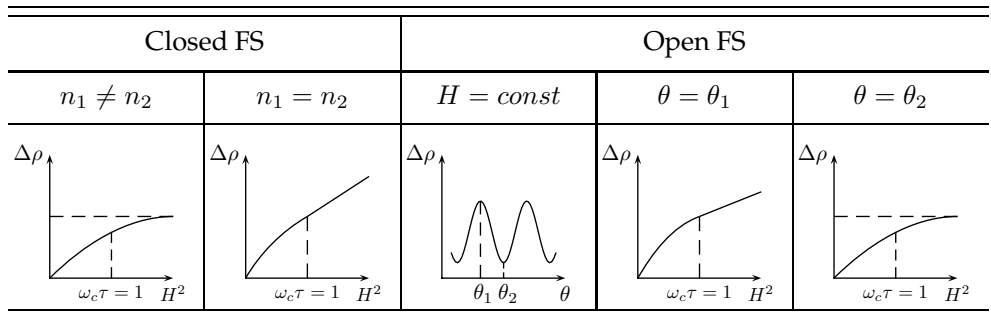} 
\caption{Behaviour of $\bm\rho=\bm\sigma^{-1}$ in metals with closed and open FS\cite{LK80}. 
(Closed) $\bm\rho$ is isotropic and saturates unless the density of electrons and holes 
coincide, in which case $\rho\sim H^2$. 
(Open) $\bm\rho$ is highly anisotropic and it shows qualitatively different behaviour 
in minima and maxima (resp. $\theta_2$ and $\theta_1$ in the picture): 
in maxima $\rho\sim H^2$, in minima it saturates ($\theta$ is the angle 
between $\bH$ and the crystallographic axis).}
\label{fig:mag}
\end{figure*}

What was clear from those early works is that, under the conditions stated above 
(and apart from the exceptional case when the density of electrons and holes are equal)
the magnetoresistance behaviour is dictated only by the topological properties
of orbits of quasi-momenta (see fig.\ref{fig:mag}): as the magnetic field $\bH$ grows, 
the magnetoresistance saturates 
isotropically to an asymptotic value if the orbits are all closed, while it grows 
quadratically with $\bH$ if there are open orbits; moreover, in this last case the 
magnetoresistance is not isotropic and the conducibility tensor $\bm\sigma$ has rank 1. 
\par
\begin{figure}
$$\vbox{\halign{\hfill#\hfill&#\hskip .15cm&\hfill#\hfill\cr
\epsfxsize=3.5cm\epsfbox[310 130 560 300]{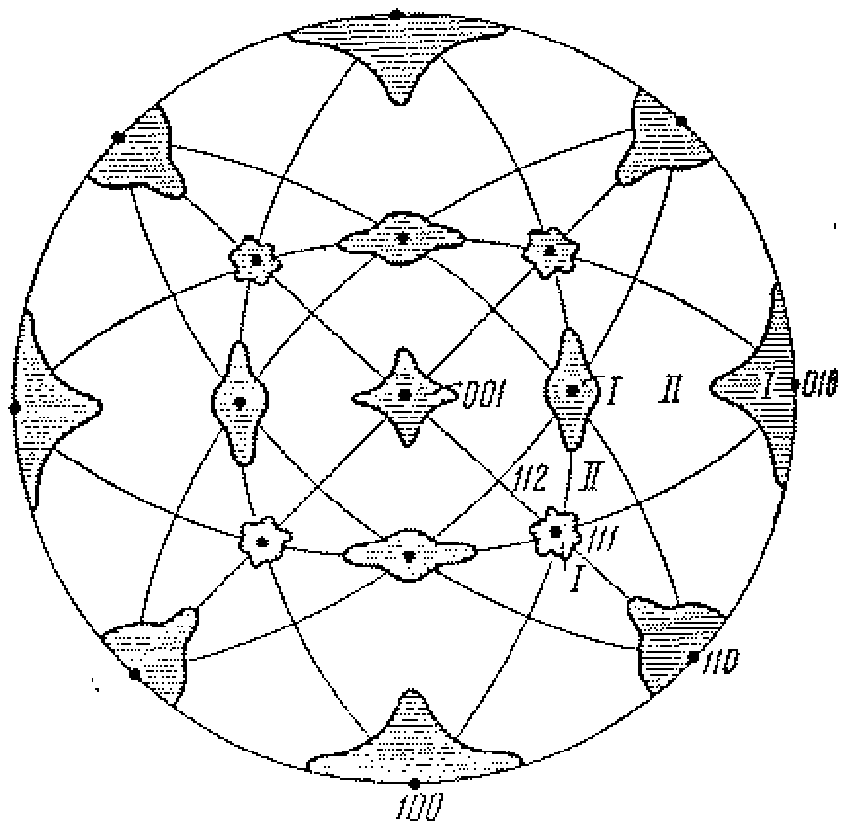}
&&
\epsfxsize=4.2cm\epsfbox[0 40 722 645]{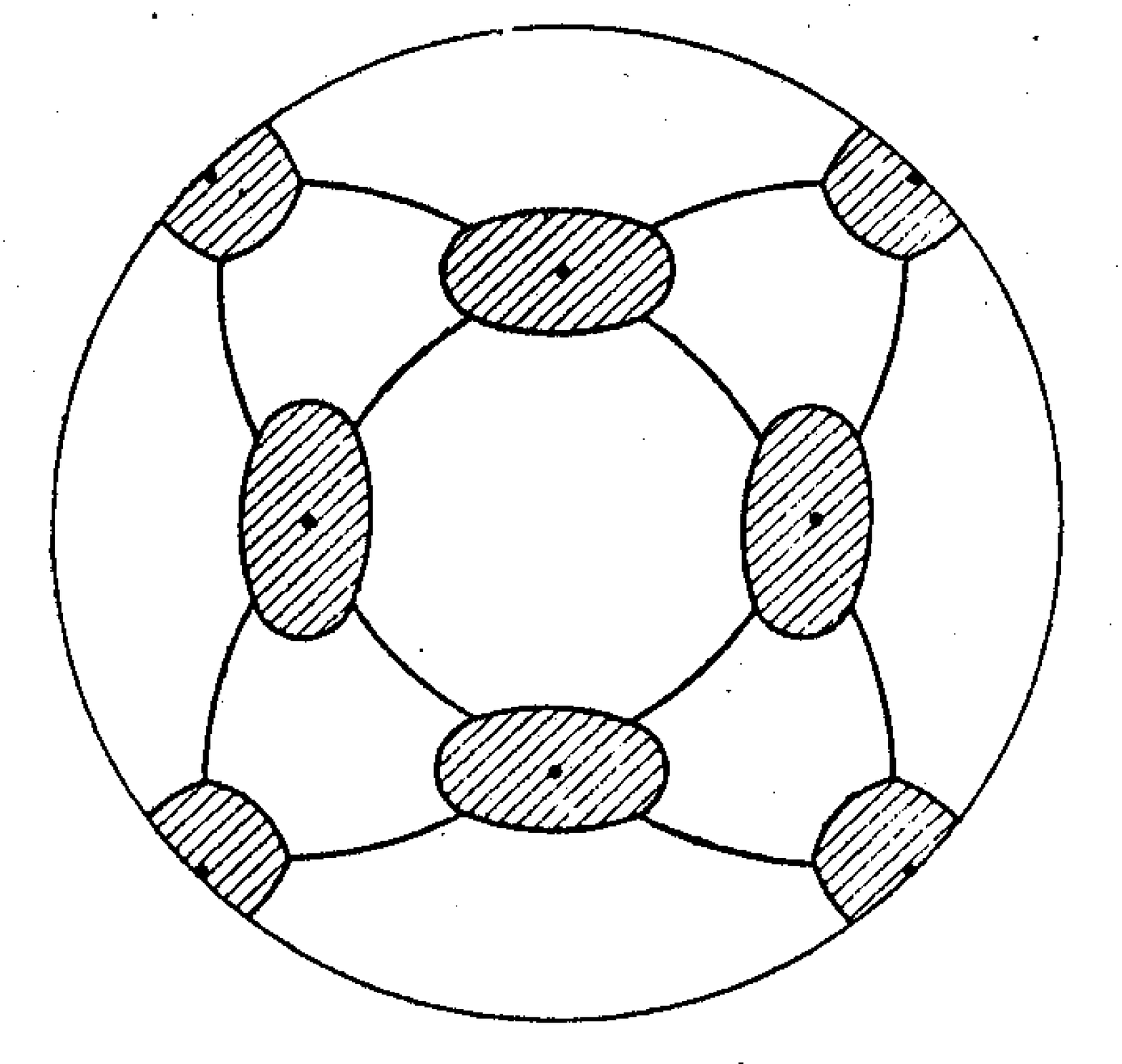}
\cr
\noalign{\vskip .4cm}	
\hskip -.1cm(a)&&\hskip -.9cm(b)\cr
}}$$
\caption{(a) SM of Gold measured by Gaidukov in 1959\cite{Ga60} 
(b) Qualitative sketch of the previous picture obtained by Lifshitz 
from his topological analysis \cite{LK60}}
\label{fig:Ga60}
\end{figure}

Many works about this effect were published between Fifties and Seventies from both the 
experimental \cite{Pip57,AG59,Ga60,AG60,AGLP61,AG62a,AG62b,AG62c,AG63,AKM64}
and theoretical \cite{LAK56,LAK57,LAK73,LP59,LP60,Cha57,Cha60,Sho60,Sho62,Pip89} point 
of view; in particular, Stereographic Maps (SM) were experimentally built by plotting 
in a stereographic projection all magnetic field directions $\bH$ in which a quadratic rise 
of $\bm\sigma$ was observed (see fig.\ref{fig:Ga60}). In those times the interest
on these magnetoresistance effects was due mainly to their utility as a tool to determine 
FS properties rather than as phenomena 
in their own right, and in particular SM maps provided information about FS topology:
e.g. clearly if $\bm\sigma$ grows quadratically for some direction of $\bH$ then the FS 
must be open, and further analysis can lead to discover the directions of the openings.
\par
Between Fifties and Seventies SM were experimentally found for about thirty metals
but, despite the theoretical efforts, no way was found to generate them with first-principles 
calculations and therefore no accurate direct verification of the Lifshitz model is 
available to date, except for the qualitative sketches by Lifshitz and Peschanskii \cite{LP60}
(see fig.\ref{fig:Ga60}); in particular it was not known till now how closely the semiclassical 
model is able to reproduce these complex experimental data and whether further
purely quanto-mechanical corrections are needed.
\par
As the magnetoresistance methods were replaced by newer and more accurate tools to
study FS, no new SM was experimentally produced and the problem was eventually abandoned; 
it was only in Nineties that the 
beautiful topological structure underlying this phenomenon, this time considered
in its own right rather than as a tool for something else, was fully discovered, 
making this way finally possible the construction of an algorithm able to reproduce 
from first-principles the experimental data about the dependence of $\bm\sigma$ 
on the direction of $\bH$ for any given Fermi Function (FF) $\cE$. 
\par
In 1982 indeed S.P. Novikov \cite{Nov82} recognized the purely topological character 
of the problem and later \cite{NM98} extracted the following generic picture from 
the work of his students A. Zorich \cite{Zor84} and I.V. Dynnikov \cite{Dyn92,Dyn97,Dyn99}: 
once a Fermi Surface (FS) is given, if open orbits arise for electrons quasi-momenta 
for some direction of the magnetic field, then the set of such directions are sorted
in some finite number of ``islands'' (e.g see fig.~\ref{fig:Ga60}); to each of these
islands it is associated a new quantum invariant $\bM$ (an irreducible Miller index 
of the lattice) that defines 
the dynamic of the semiclassical system in the following way: each open orbit 
corresponding to a $\bH$ that belongs to an island labeled by $\bM$ is a finite 
deformation of a straight line parallel to the vector product $\bH\times\bM$. 
All $\bH$ that do not fall in any of these islands give rise only to closed orbits, 
except for a negligible set of exceptional directions that we will disregard here.
\par
In this article we present the method, suggested to us by I. Dynnikov, that we implemented, 
based on this picture, to detect the directions of the magnetic field for which open 
orbits of quasi-momenta appear. This algorithm allows us to predict the SM of a metal
according to the Lifshitz model, namely to determine the part of the ``islands'' 
that the semiclassical approximation is sufficient to detect. 
As an application, we first study the SM of a very rich and simple toy model and then
we generate SM for Au and Ag and compare our results with the ones obtained 
experimentally more than forty years ago by Gaidukov (and never repeated since then); 
this represents the first rigorous check 
of Gaidukov's results from first-principles and shows that, even though the SM were measured 
just at the lower threshold for the semiclassical approximation to hold, the Lifshitz model 
is able to reproduce rather accurately the experimental data.
\section{Overview of the Lifshitz model}
A great deal of energies have been invested in Fifties ans Sixties in the study of
magnetoresistance in a metal. After the discovery by Kapitza \cite{Kap29} of a linear increase
of the resistivity with a magnetic field in a number of metals, it had be shown
by Justi \cite{JS37} that metals could be divided in two categories: for the first one the 
resistivity saturates with the increase of the magnetic field (e.g. in Cu, Na, Al)
while for the second one it grows quadratically (e.g. in noble metals).
\par
It was Peierls \cite{Pei31} the first to recognize that this anomalous behaviour was due 
to the departures from the free-electron model, but it was especially thanks to the 
theoretical works of I.M. Lifshitz and his school \cite{LAK56,LP59,AGLP61}
and to the experimental results of Alekseevskii and Gaidukov 
\cite{Ga60,AG59,AG60,AGLP61,AG62a,AG62b,AG62c,AG63,AKM64} that a systematic and thorough 
study of this phenomenon was carried on and fully revealed its critical
dependence on the topology of the FS. In particular, Lifshitz was the first to 
study the system in its full generality, making no assumption on the form of the FF.
\par
Let us review quickly the mathematics of the model to show the role topology has in it 
\cite{AM76,NM03}.
Since no analytical method to solve exactly the Schrodinger equation under a generic periodic 
potential is known, we will make, as usual, a first approximation introducing the semiclassical 
model, i.e. we neglect the electron-electron interaction and consider the motion of a 
single electron in an infinite crystal.
According to the semiclassical approximation, the Schrodinger equation 
for a single electron in a 
three-dimensional lattice $\Gamma$ with a $\Gamma$-invariant potential in presence
of an electric field $\bE$ and a magnetic field $\bH$ reduces to the following 
(semi-)classical equations of motion:
$$
\vbox{\halign{$#$\hfill&$#$\hfill\cr
     \dot\bq&=v_n(\bp)=\frac{\partial \cE_n(\bp)}{\partial\bp}\cr
     \dot\bp&=-e\left[\bE+\frac{1}{c}v_n(\bp)\times\bH\right]\cr
}}
$$
where $\cE_n(\bp)$ is the energy function for the electron occupying the band $n$. 
\par
Physical constraints limit the range of the magnetic field:
electrons become aware of the FS topology only if the mean free path is long 
enough to traverse considerable portions of it, that requires $\omega_c\tau\gg1$
(and so a field $H\geq10T$), a pure crystal and very low temperatures; on the
other side, to avoid magnetic breakdown we also must have $H\leq10^3T$.
\par
The semiclassical approximation makes the problem look at first sight as a standard 
classical mechanics system but it turns out that it is instead deeply different from 
all of them. The difference is not in
the analytical expression of the equations, that is evidently the same, but rather 
in the {\sl topology} of the phase space: in classical mechanics indeed the momenta 
belongs always to a linear space, no matter how complicated the base space is; 
here instead the base space is topologically trivial (the whole three-space) but the 
momenta are triply periodic by the Bloch theorem, since we identify all momenta that 
differ by a vector of the dual lattice $\Gamma^*$.
In more rigorous terms, while in classical mechanics momenta
would belong to the linear space $\tR$, in this case they are defined only modulo a
vector of the reciprocal lattice and therefore they belong to the first Brillouin zone,
i.e., in other words, to the three-torus $\tT=\tR/\Gamma^*$.
\par
This difference is {\sl essential} because it is exactly what brings topology 
(in this case ``periodic topology'') in play: the 
three-torus, differently from the three-space, has a non-trivial topology and its presence 
has a strong influence on the dynamics of the system.
Indeed, if both $\bE$ and $\bH$ are constant, the pair of (systems of) equations
de-couples and the problem reduces to study the orbits of the quasi-momenta in 
the first Brillouin zone (with the obvious boundary conditions dictated by
the periodicity) under the equation:
$$\dot\bp=-e\left[\bE+\frac{1}{c}\frac{\partial\cE(\bp)}{\partial p}\times\bH\right]$$
\par
As often happens, topological effects are due to the magnetic field rather than to the 
electric one and therefore we can safely put $\bE=\bm0$ and $e=c=1$ and we are 
finally left with the equation:
$$\dot\bp=-\frac{\partial\cE(\bp)}{\partial p}\times\bH=\{\bp,\cE(\bp)\}_{_{\bH}}$$
where $\{,\}_{_{\bH}}$ is the so-called ``magnetic bracket'':
$$\{p_\alpha,p_\beta\}_{_{\bH}}=\epsilon_{\alpha\beta\gamma} H^\gamma$$
It is well known that this system is over-integrable since it has two integrals of motion, 
namely the Hamiltonian $\cE$ and the component $\bp\cdot\bH$ of the quasi-momentum 
in the magnetic field 
direction; nevertheless it was not fully understood until Eighties the 
relevance of the fact that the latter integral is not a well defined {\sl single-}valued 
function in $\tT$ but rather a {\sl multi}-valued function, exactly in the same way the 
angle $\theta$ is just a multivalued function in the circle $\bS^1$.
\par
Let us point out that not even the most elementary systems with multivalued first integrals 
have been object of study till recent years by the dynamical systems community, probably because 
no such system arises naturally from classical mechanics problems.
This appearance of a multivalued first integral, whose occurrence is due to the non-trivial 
topology of $\tT$ and could not arise in a topologically trivial space like $\tR$, 
it's enough to transform an otherwise trivial dynamical system in an extremely rich one. 
For example, cutting level surfaces of $\cE$ with level surfaces of another well-defined 
function on $\tT$ would lead only to closed orbits in the Brillouin zone, while in our
case open orbits do generically arise, orbits that may in principle fill the whole
FS in the same way a straight line with 3-irrational slope would fill the whole Brillouin
zone.
%
%
\par
The topology of the orbits, namely whether they are open or closed (in the repeated zone 
scheme), is not just a mathematical curiosity:
it has been first showed by Lifshitz indeed that the cause of the quadratic growth of the
magnetoresistance is exactly the presence of open orbits.
A rigorous deduction of this fact can be made using the Boltzman transport equation \cite{Abr88}
but simpler justifications of it can be made \cite{LK80} at a phenomenological level 
based on the Einstein relation $\sigma\simeq\cD e^2n/\cE_F$, where $\cD$ is the diffusion
coefficient and $\cE_F$ the Fermi Energy. 
When all orbits are closed indeed the diffusion in the plane perpendicular to $\bH$
consists in jumps by an amount of the order of the cyclotron radius $r_H=cp_F/eH$ 
with frequency $\sim 1/\tau$, so that $\cD\simeq lv\simeq r^2_H/\tau$ and
$\bm\sigma\simeq\sigma_0/(\omega_c\tau)^2$; now suppose instead that quasi-momenta open 
orbits appear instead, say in the $x$ direction: then the electrons will move along the
$y$ direction, since the orbits in the real space are rotated by $\pi/2$ with respect to
orbits in the momentum space, so that in the $x$ direction $\cD$ has more or less 
the same $\bH$ dependence found for the closed orbits but in the orthogonal direction 
electrons move as free particles, namely $\cD_{yy}\simeq v^2_F\tau$, and therefore in 
this case $\bm\sigma$ is not isotropic anymore and
$\sigma_{xx}\simeq\sigma_0/(\omega_c\tau)^2$, $\sigma_{yy}\simeq\sigma_0$.
\par
The theoretical results by Lifshitz and his school urged Alekseevskii and Gaidukov 
to start conducting careful experiments that turned out to be in perfect agreement
with the model. The most interesting experimental result for us is the 
stereographic projection of the special direction for the magnetic field
(Stereographic Map), namely the map on the unitary disc that shows for which 
directions the magnetoresistance grows quadratically.
Such map indeed, in the semiclassical approximation, depends solely on the orbits 
topology and therefore it is totally determined once a Fermi Surface is given.
A great effort has been spent to find some kind of algorithm able to produce
this map from a generic FS \cite{LP59,LP60,Cha60}, but the topological tools to solve 
the problem were not known to physicists at that time and eventually the problem 
was left unsolved.
\section{Overview of the recent topological results}
\label{sec:top}
The interest in this problem revived in 1982 when S.P. Novikov found out that it
was a perfectly suited case where to apply his newly introduced Morse-Novikov theory 
\cite{Nov82}, namely the study of the topology of level surfaces of multivalued functions.
\par
Fundamental results were found in Eighties and Nineties by his pupils A.V. Zorich
and I.A. Dynnikov and from them Novikov later extracted the following picture:
once a ``complicated enough'' Fermi Function is given (we will clarify this concept below), 
a fractal is determined on the set of the stereographic 
projections of the magnetic field directions; the fractal consists of smooth
polygons (islands, or ``stability zones'') that contain all possible lattice directions and 
generically meet each other in a finite number of points; every such polygon is labeled 
by a Miller index $\bM$ and moreover to every magnetic field direction are associated 
two values of the energy $e_{m,M}(\bH)$. 
\par
\begin{figure}
\epsfxsize=7cm\epsfbox{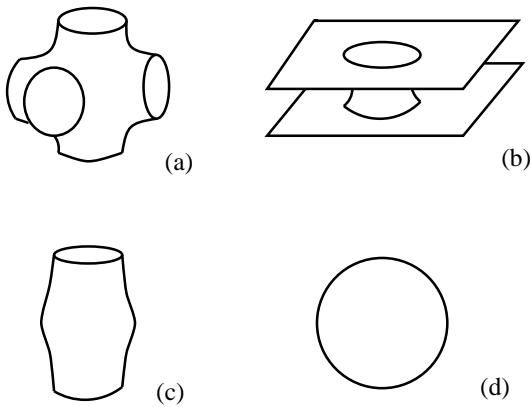}
\caption{FS with rank  3 (a), 2 (b), 1 (c) and 0 (d).}
\label{fig:rank}
\end{figure}
The meaning of this picture is the following. Let $\cE$ and $E_F$ be the Fermi Function 
and Fermi Energy of a metal and suppose that we want to know the asymptotic behavior of 
trajectories of quasi-momenta for some magnetic field $\bH$ whose stereographic 
projection lies inside a polygon labeled by the Miller index $\bM$: 
the answer is that
if $E_F$ lies between $e_m(\bH)$ an $e_M(\bH)$ then open orbits exist and moreover
they are all finite deformations of a straight line with direction $\bH \times \bM$;
if $E_F$ is instead smaller than $e_m$ or bigger than $e_M$ then all orbits are closed.
\par
The directions for which  $e_m(\bH)=e_M(\bH)$ are non-generic and Novikov conjectured
that they form a set of measure zero in the disc.  The orbits of quasi-momenta 
induced by magnetic fields with those directions are much more complicated than the 
generic ones and for each such fixed direction there is only one energy value for which 
open orbits appear, while at every other energy all orbits are closed. 
Nevertheless this case may be relevant since it may explain deviations from the quadratic 
growth law detected in several metals for special directions of the magnetic field and 
it is currently under investigation by Novikov himself and Ya. Maltsev \cite{NM03}.
\par
Below we review in some detail the topological ideas on which the method is based.
The first important fact is 
that \cite{NDF89} every FS is topologically equivalent to a sphere with some 
finite number of handles attached (such number is called the {\sl genus} of the FS).
It is an intuitive fact that we need a genus bigger than zero to have an open FS, but 
this condition is not sufficient: it is enough to think to a sphere in the center of the
first Brillouin zone and to add to it handles without ever touching the boundary of the
zone.
\par
The concept that detects whether a FS is open or not is its 
{\sl topological rank}, namely the complement to three of the biggest number of 
``linearly independent'' pairs of planes (i.e. planes whose perpendiculars are 
linearly independent) that can enclose the FS. Examples of FS of rank from three 
to zero are shown in fig.\ref{fig:rank}: 
for example a sphere can be enclosed between three linearly independent pairs of planes, 
while a cylinder can be enclosed only between at most two of such pairs and so on. 
The case of interest for the phenomenon in study is the rank-3 one, since in the other 
ones either all orbits are closed (rank zero) or generically closed (rank one) or open 
but with an obvious asymptotic direction (rank two).
\par
\begin{figure}[b]
\pspicture(0,3.5)
\epsfxsize=6cm\epsfbox{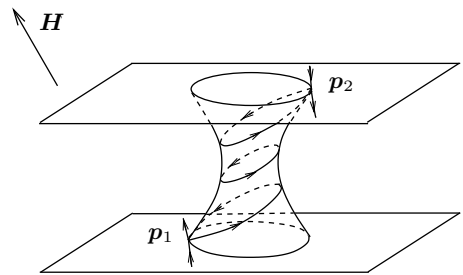}
\pscircle[fillstyle=solid,fillcolor=white,linecolor=white](-5.5,3.3){.2}
\rput(-5.45,3.35){$\bH$}
\rput(-1.6,2.5){$\bp_2$}
\rput(-4,.5){$\bp_1$}
\endpspicture
\caption{A genus-2 rank-2 surface. For this choice of $\bH$ closed loops are cut on the
cylinder and therefore the critical saddles through $p_1$ and $p_2$ are half-opened 
(fig.\ref{fig:Zor}(ii)) and open orbits arise}
\label{fig:cyl}
\end{figure}
\begin{figure}
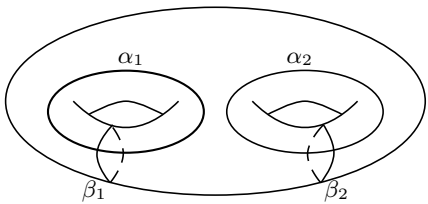

\psset{unit=.7cm,linewidth=.6pt}
\pspicture(0,0)(6,4)
%
\psellipse(3,2)(4,1.8)
\parabola(.3,2)(1.3,1.5)
\pscurve(.6,1.75)(1.3,2)(2,1.75)
\rput(1.4,2.8){$\alpha_1$}
\psellipse[linewidth=.8pt](1.3,1.8)(1.5,.8)
\rput(.7,.3){$\beta_1$}
\pscurve[curvature=1 1 0](1.05,1.55)(.75,1)(1,.45)
\pscurve[curvature=1 1 0,linestyle=dashed](1.05,1.5)(1.25,1)(1,.45)
\parabola(3.7,2)(4.7,1.5)
\pscurve(4,1.75)(4.7,2)(5.4,1.75)
\rput(4.6,2.8){$\alpha_2$}
\psellipse(4.7,1.8)(1.5,.8)
\rput(5.3,.3){$\beta_2$}
\pscurve[curvature=1 1 0](5.05,1.55)(5.25,1)(5,.45)
\pscurve[curvature=1 1 0,linestyle=dashed](5.05,1.55)(4.75,1)(5,.45)
\endpspicture
\caption{Possible choice of canonical loops in a genus-2 surface. Any other 
non-trivial loop (i.e. a loop that cannot be continously shrunk to a point)
drawn on the surface cuts at least one of the base loops, 
i.e. it is ``linearly dependent'' by them.}
\label{fig:hom}
\end{figure}

Despite its triviality, genus-2 rank-2 case turns out to be paradigmatic and 
it is worth describing it in detail. Examples of such surface are shown in fig.\ref{fig:hom},
\ref{fig:rank}(b),\ref{fig:cyl}, namely a pair of parallel rectangles joined 
by a cylinder (incidentally, this is exactly the topology of the FS of Tin).
The normal to the rectangles is a lattice direction that we will call $\bL$.
\par
\begin{figure*}
$$\vbox{\halign{\hskip2.cm#\hskip 2.5cm&\hfill#\hfill&\hskip 1.5cm#&\hfill#\hfill\cr
&
\psset{unit=.45cm}
\pspicture(0,0)(8,8)
\psline(2,7.5)(0,7.5)(-2,5.5)(8.5,5.5)(10,7.5)(7.57,7.5)
\psline[linestyle=dashed,dash=3pt 2pt](2,7.5)(3,7.5)
\psline[linestyle=dashed,dash=3pt 2pt](7.57,7.5)(6.43,7.5)
\psline(6.43,7.5)(3,7.5)
\psline(.05,4)(-1,4)(-3,2)(8.5,2)(10,4)(6.1,4)
\psline[linestyle=dashed,dash=3pt 2pt](.05,4)(.95,4)
\psline(.95,4)(3.57,4)
\psline[linestyle=dashed,dash=3pt 2pt](3.57,4)(4.43,4)
\psline(4.43,4)(4.9,4)
\psline[linestyle=dashed,dash=3pt 2pt](4.9,4)(6.1,4)

\psellipse(.5,6.2)(.45,.2)
\parametricplot[linestyle=dashed,dash=3pt 2pt]{0}{180}{t cos .45 mul .5 add t sin .2 mul 2.8 add}
\parametricplot{180}{360}{t cos .45 mul .5 add t sin .2 mul 2.8 add}
\psline[linestyle=dashed,dash=3pt 2pt](.05,6.2)(.05,5.5)\psline(.05,5.5)(.05,2.8)
\psline[linestyle=dashed,dash=3pt 2pt](.95,6.2)(.95,5.5)\psline(.95,5.5)(.95,2.8)

\parametricplot[linestyle=dashed,dash=3pt 2pt]{0}{180}{t cos .5 mul 2.5 add t sin .16 mul 7 add}
\parametricplot{180}{360}{t cos .5 mul 2.5 add t sin .16 mul 7 add}
\psline(2,7)(2,8)\psline[linestyle=dashed,dash=3pt 2pt](2,3.5)(2,2)\psline(2,2)(2,1)
\psellipse(2.5,3.5)(.5,.16)
\psline(3,7)(3,8)\psline[linestyle=dashed,dash=3pt 2pt](3,3.5)(3,2)\psline(3,2)(3,1)

\psellipse(4,6)(.43,.15)
\parametricplot[linestyle=dashed,dash=3pt 2pt]{0}{180}{t cos .43 mul 4 add t sin .15 mul 2.5 add}
\parametricplot{180}{360}{t cos .43 mul 4 add t sin .15 mul 2.5 add}
\psline[linestyle=dashed,dash=3pt 2pt](3.57,6)(3.57,5.5)\psline(3.57,2.5)(3.57,5.5)
\psline[linestyle=dashed,dash=3pt 2pt](4.43,6)(4.43,5.5)\psline(4.43,5.5)(4.43,2.5)

\psellipse(5.5,7)(.6,.2)
\parametricplot[linestyle=dashed,dash=3pt 2pt]{0}{180}{t cos .6 mul 5.5 add t sin .2 mul 3.5 add}
\parametricplot{180}{360}{t cos .6 mul 5.5 add t sin .2 mul 3.5 add}
\psline[linestyle=dashed,dash=3pt 2pt](4.9,7)(4.9,5.5)\psline(4.9,5.5)(4.9,3.5)
\psline[linestyle=dashed,dash=3pt 2pt](6.1,7)(6.1,5.5)\psline(6.1,5.5)(6.1,3.5)

\parametricplot[linestyle=dashed,dash=3pt 2pt]{0}{180}{t cos .57 mul 7 add t sin .2 mul 6.5 add}
\parametricplot{180}{360}{t cos .57 mul 7 add t sin .2 mul 6.5 add}
\psline(6.43,6.5)(6.43,8)\psline(7.57,6.5)(7.57,8)
\psellipse(7,3)(.57,.2)
\psline[linestyle=dashed,dash=3pt 2pt](6.43,3)(6.43,2)\psline(6.43,2)(6.43,1)
\psline[linestyle=dashed,dash=3pt 2pt](7.57,3)(7.57,2)\psline(7.57,2)(7.57,1)

\endpspicture
&&
\pspicture(4,4)
\epsfxsize=5cm\epsfbox{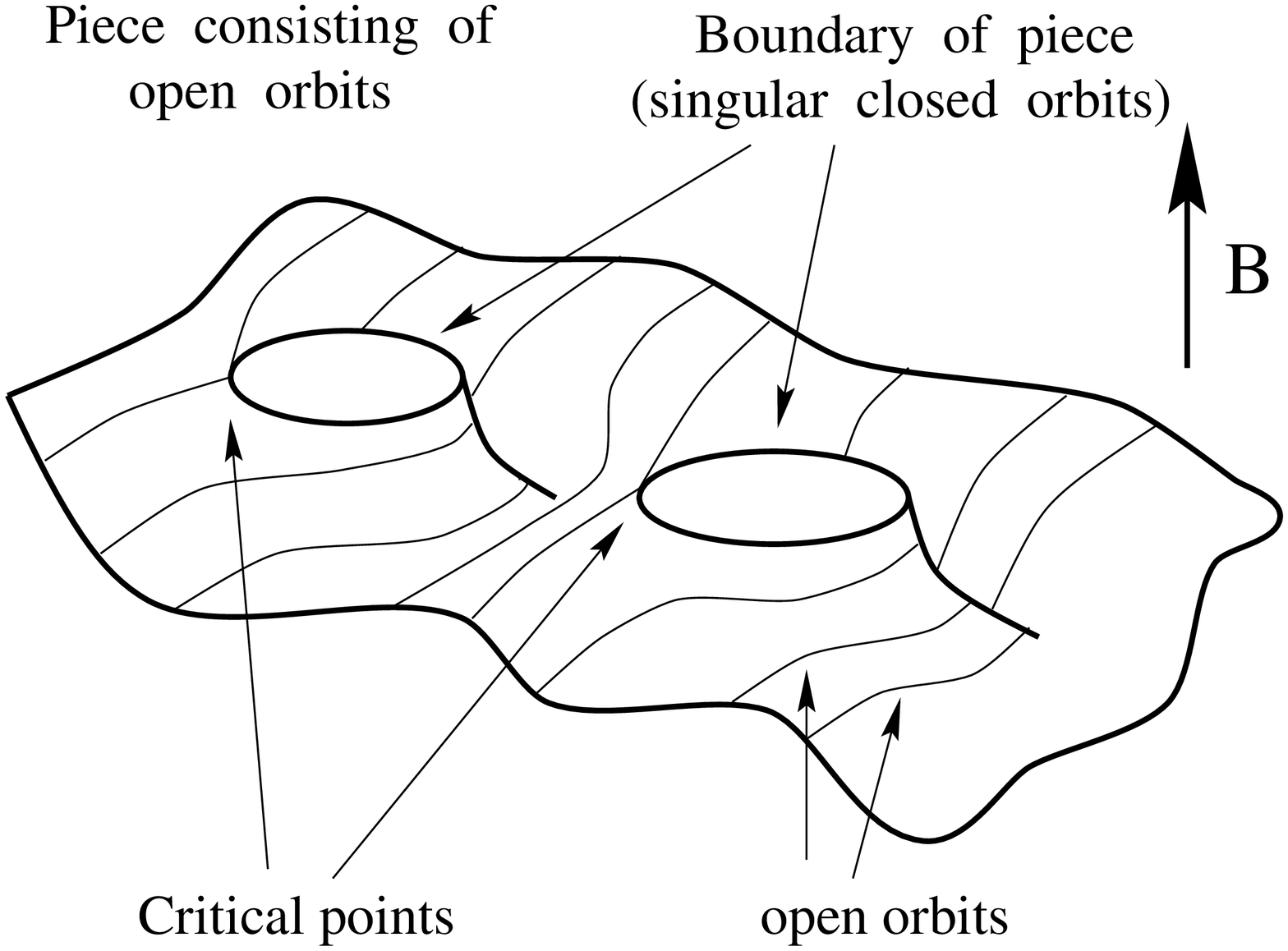}
\pscircle[fillstyle=solid,fillcolor=white,linecolor=white](-.1,2.7){.2}
\rput(-.1,2.7){$\bH$}
\endpspicture
\cr
}
}$$
\caption{(a) Generic picture in case open orbits exist (here $g=8$); 
(b) warped plane filled by open orbits (the two discs shown are bases for
two cylinders of closed orbits\cite{NM03} (not shown in fig.)}
\label{fig:wp}
\end{figure*}
%
Suppose first that $\bH$ induces closed orbits on the cylinder (as in fig.\ref{fig:cyl}): 
then $\bH$ induces also open orbits, since any orbit that does not lie entirely on the 
cylinder will be bound to stay on only one of the two plane sheets and therefore will be 
open (it has no way to turn back). 
Moreover, the surface is clearly enclosed between a pair of parallel lattice planes and 
therefore the orbits lie entirely in a finite width strip and, as Dynnikov \cite{Dyn92} 
showed, pass through it; the Miller index common to the two lattice planes is clearly
$\bL$ and it is the quantum invariant associated to $\bH$.
Suppose now that there exist a plane perpendicular to $\bH$ that intersects both bases
of the cylinder, so that no closed orbit is cut on the cylinder by any plane parallel to it: 
then, assuming $\bH$ fully irrational to simplify the discussion, every orbit will be closed.
Indeed take any point on the FS and follow its orbit in one of the directions: since $\bH$
is fully irrational, at a certain point the plane generating the orbit will cut a copy of
the cylinder (in the extended zone picture) on both bases, i.e. the orbit will turn back;
since exactly the same happens in the other direction, the orbit is closed.
\par
Our analysis shows that the SM corresponding to such a FS, as it has been obtained
experimentally for Tin \cite{AGLP61}, contains a single island, whose boundary is given 
by the set of directions of planes that are tangent to both bases in diametrally opposite
points. For example, if the topological cylinder is actually a right circular cylinder
of radius $r$ and height $h$, the island is a circle with center in the $\bM$ direction
and radius $h/r$.
The Miller index associated to this island is of course the same $\bM$
above and the direction of open orbits is given by the vector product $\bM\times\bH$.  
\par
Let us come now to the rank-3 case.
A fundamental result by Zorich \cite{Zor84} is that, no matter how complicated 
the FS is, the one above is the generic behaviour for all directions ``close enough''
to rational, namely the open orbits, when they arise, lie on components of the 
FS that are exactly ``warped planes'' separated by cylinders of closed orbits. 
For each fixed direction, all such planes are parallel to each other and are in even number.
The number of pairs is bound from above by $g/2$, where $g$ is the genus of the surface; the 
number of cylinders is bounded from below by $g-1$ (see fig.\ref{fig:wp}).
The difference with the rank-$2$ case is that now the surface may be split
in many different ways in ``warped planes'' and cylinders (even in infinitely many
ways in special cases), so that more than one zone can appear (possibly each one 
with a different Miller index).
\par
\begin{figure}[b]
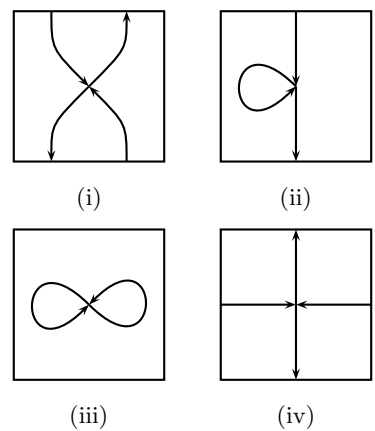

$$\vbox{\halign{\hfill#\hfill&#\hskip.75cm&\hfill#\hfill\cr
\psset{unit=.5cm}
\pspicture(0,0)(4,4)
\pspolygon(0,0)(4,0)(4,4)(0,4)
\psbezier{->}(2,2)(3,3)(3,3)(3,4)
\psbezier{<-}(2,2)(1,3)(1,3)(1,4)
\psbezier{<-}(2,2)(3,1)(3,1)(3,0)
\psbezier{->}(2,2)(1,1)(1,1)(1,0)
\endpspicture
&&
\psset{unit=.5cm}
\pspicture(0,0)(4,4)
\pspolygon(0,0)(4,0)(4,4)(0,4)
\psline{<-}(2,2)(2,4)
\psline{->}(2,2)(2,0)
\psbezier{->}(2,2)(0,4)(0,0)(2,2)
\endpspicture
\cr
\noalign{\vskip 6pt}
(i)&&(ii)\cr
\noalign{\vskip 6pt}
\psset{unit=.5cm}
\pspicture(0,0)(4,4)
\pspolygon(0,0)(4,0)(4,4)(0,4)
\psbezier{->}(2,2)(0,4)(0,0)(2,2)
\psbezier{->}(2,2)(4,0)(4,4)(2,2)
\endpspicture
&&
\psset{unit=.5cm}
\pspicture(0,0)(4,4)
\pspolygon(0,0)(4,0)(4,4)(0,4)
\psline{->}(2,2)(2,4)
\psline{->}(2,2)(2,0)
\psline{<-}(2,2)(0,2)
\psline{<-}(2,2)(4,2)
\endpspicture
\cr
\noalign{\vskip 6pt}
(iii)&&(iv)\cr
}}$$
\caption{Possible kind of open saddles for rational $\bH$: $(i)$ Fully open; 
$(ii)$ Half-open; $(iii)$ Fully closed; $(iv)$ Fully open (impossible because of the boundary
conditions). Only $(ii)$ plays a role in the generic case since $(i)$ arises only for 
rational directions of $\bH$.}
\label{fig:Zor}
\end{figure}
The key-point here is the strong constraint determined by periodicity to the
type of critical point that can be met in the sections of the FS by planes.
Consider the simple case of $\bH=(0,0,1)$: assuming that there are open orbits and
that we are not in a degenerate case, all critical points will appear at different 
levels and they will be of one of the types shown in fig.\ref{fig:Zor}; since we are 
interested in open orbits, case $(iii)$ is irrelevant for us and case 
$(iv)$ is forbidden by the boundary onditions. 
Let us now follow one of the open orbits: if it never meets any other curve, then after 
a period it meets itself again, meaning that the whole FS is just a single warped plane 
and everything is trivial (it is a genus-1 surface with rank 1 or 2); 
if it meets another orbit, then either it will be another open 
orbit (case (i)) and the two will annihilate each other (or, from another 
point of view, the open orbit will hit a closed one and bounce back), or it will be 
a closed orbit (case (ii)) and so the open one will simply engulf it and go on till it will
meet again itself.
There are only two possible outcomes for this process: either
a warped cylinder (genus-1, rank-1) or a warped plane (genus-1, rank-2), in each case 
with some finite number of plane holes that are the basis of the cylinders of closed 
orbits that separate them from each other (see fig.\ref{fig:wp}).
\par
The reasonment above holds for any $\bH$ pointing to a lattice direction, since we can 
bring every such $\bH$ to $(0,0,1)$ with a coordinate change that leaves the lattice invariant, 
but it is impossible to extend it to the case of ``irrational'' $\bH$ 
(i.e. not directed along a lattice directions), since in that case the orbits are not periodic 
anymore. Nevertheless, continuity helps us there:
indeed the heights of the cylinders of closed orbits depend (at least) continuously on 
the direction $\bH$ and therefore the cylinders will survive small perturbations
and open orbits will still appear and will be bound to lie on small deformations of the same 
warped planes.
\par
The Zorich observation reveals the existence of a quantum invariant of the system that 
had previously been missed and, at the same time, explains why magnetic field directions in
SM are sorted in islands: small deformations of warped planes preserve open orbits and lead 
to the same Miller index $\bL$, so there is a whole island around each rational $\bH$ giving 
rise to open orbits and the same $\bL$ is associated to all directions inside the island.
From the
construction it is clear that the open orbits have direction given by $\bH\times\bL$,
so that experimentally it is enough to determine the direction of open orbits for two 
directions of the island to determine uniquely $\bL$.
\par
%
This discovery though does not solve completely the problem, since the set 
of directions ``close enough'' to all rational directions may not, in principle, 
cover the whole the disc. In Nineties I. Dynnikov showed \cite{Dyn99} that the
Zorich picture does represent the generic behaviour of the system in the following
sense: given a Fermi Function $\cE(\bp)$ and a direction $\bH$, open orbits arise 
only for a closed interval of energies $[e_m(\bH),e_M(\bH)]$ and the structure of
open orbits is the one described above when $e_m(\bH)=e_M(\bH)$; in particular this
means that, for every possible direction of $\bH$, there is at least one value $e$ of
the energy such that $\bH$ induces open orbits on $\cE(\bp)=e$. 
\par
Dynnikov's results also lead to a conclusion rather interesting and unexpected: 
SM taken at different energies are compatible (i.e. no more than one $\bL$ is 
associated to every $\bH$) and when we plot them all at the same time, building some 
kind of ``all energies'' SM, there are only two possible cases: either just one zone 
is present
\footnote{e.g. this happens for FF whose level sets are only spheres and Tin-like surfaces}, 
and that zone fills the whole disc, or infinitely many zones appear,
distributed in a fractal-like way (see fig.\ref{fig:g3frac}).
Even though we cannot freely change the Fermi Energy of a metal, these fractals
may in principle play a role in Physics because, for some special class of 
functions, there is an energy level at which the standard SM coincides exactly
with the ``global SM'' (an example is provided in sec.\ref{sec:g3}).
\section{The method}
The topological picture discovered by Zorich makes relatively simple to build 
an algorithm able to reconstruct the magnetoresistance map from any given FS.
\par
Indeed, from what we said in the section above, it is clear that to investigate the
topology of the orbits there is no need to study their asymptotics but rather it is 
enough to evaluate the Miller index associated to every magnetic field direction.
Moreover, because of the Zorich result, it is enough to study what happens in the 
case of rational directions; this changes {\sl qualitatively} the nature of the numerical 
problem we have to deal with, since the FS sections by a Miller plane are 
{\sl periodic} and therefore, in principle, the {\sl whole} orbits can be numerically 
evaluated with the desired precision.
\par
To evaluate the Miller index we need to use elementary homology properties of loops on
surfaces. It is a well known fact in topology that the algebraic sum of the intersections 
between two loops ({\sl intersection number}) on closed compact surfaces is a homology 
invariant, i.e. it does not change if  
a loop is deformed with continuity or if it is replaced with one that forms with it the 
boundary of some surface. Moreover, if we agree to consider as the same loop all
loops that are homologous to each other, we can give a $\bZ$-linear structure to this set 
that turns out to be isomorphic to $\bZ^{2g}$, i.e. it is (freely) generated by $2g$ 
(classes of) loops (see fig.\ref{fig:hom}).
In this setting, the intersection number becomes a bilinear antisymmetric non-degenerate 
form $\langle,\rangle$, namely a symplectic structure on $\bZ^{2g}$, and in any canonical base 
$(\alpha_i,\beta_j)_{_{i,j=1,..,g}}$ we have that 
$$<\alpha_i,\beta_j>=1\,,\,\,<\alpha_i,\alpha_j>=<\beta_i,\beta_j>=0\,\,.$$
\par
The key observation is that the loops that lie on the 
``warped planes'' of open orbits have all intersection number zero with the closed loops
living on the cylinders, since in that case there is no intersection at all.
To evaluate the desired Miller index therefore it is enough to find out the homology
classes of all loops that have zero intersection number with the loops that form
the closed cylinder and finally find out the lattice direction that these loops 
have in the reciprocal lattice $\Gamma^\star$: since all of them lie on ``warped planes'' 
homologous to the same lattice plane, the result will be a two-dimensional sublattice of 
$\Gamma^\star$ that will automatically give the Miller index we looked for.
\par
We implemented this method in the following way: once a FS $M$ and a direction $\bH$ 
are given, first of all we identify somehow a base (better if canonical) for the homology 
loops on $M$ and provide methods to evaluate the intersection number with 
respect to them; then we determine all saddle points of type $(i)$ and $(ii)$ that $\bH$ 
induces on $M$ and evaluate the intersection number of the closed loops of each of
these saddles with all loops of the base above. Once all these data have been collected
(and after many compatibility checks have been made) we identify the space of non-trivial 
loops that do not intersect them,
namely we find a base for the space of all loops of $M$ that lie on the warped planes 
\footnote{they are closed loops on $M$ when we look at them in the reduced zone scheme but, 
once seen in the repeated zone scheme, they are open}.
Finally, we evaluate the directions of those loops in the reciprocal lattice $\Gamma^*$, 
that will give us a set of directions identifying a 2-dimensional sublattice of
$\Gamma^*$ that in turn will give us the desired Miller index.
\par
\begin{figure}
$$\vbox{
\halign{\hfill#\hfill&#\hskip .5cm&\hfill#\hfill\cr
\epsfxsize=3.75cm\epsfbox[50 50 390 290]{fig6a.eps}
&&
\epsfxsize=3.75cm\epsfbox{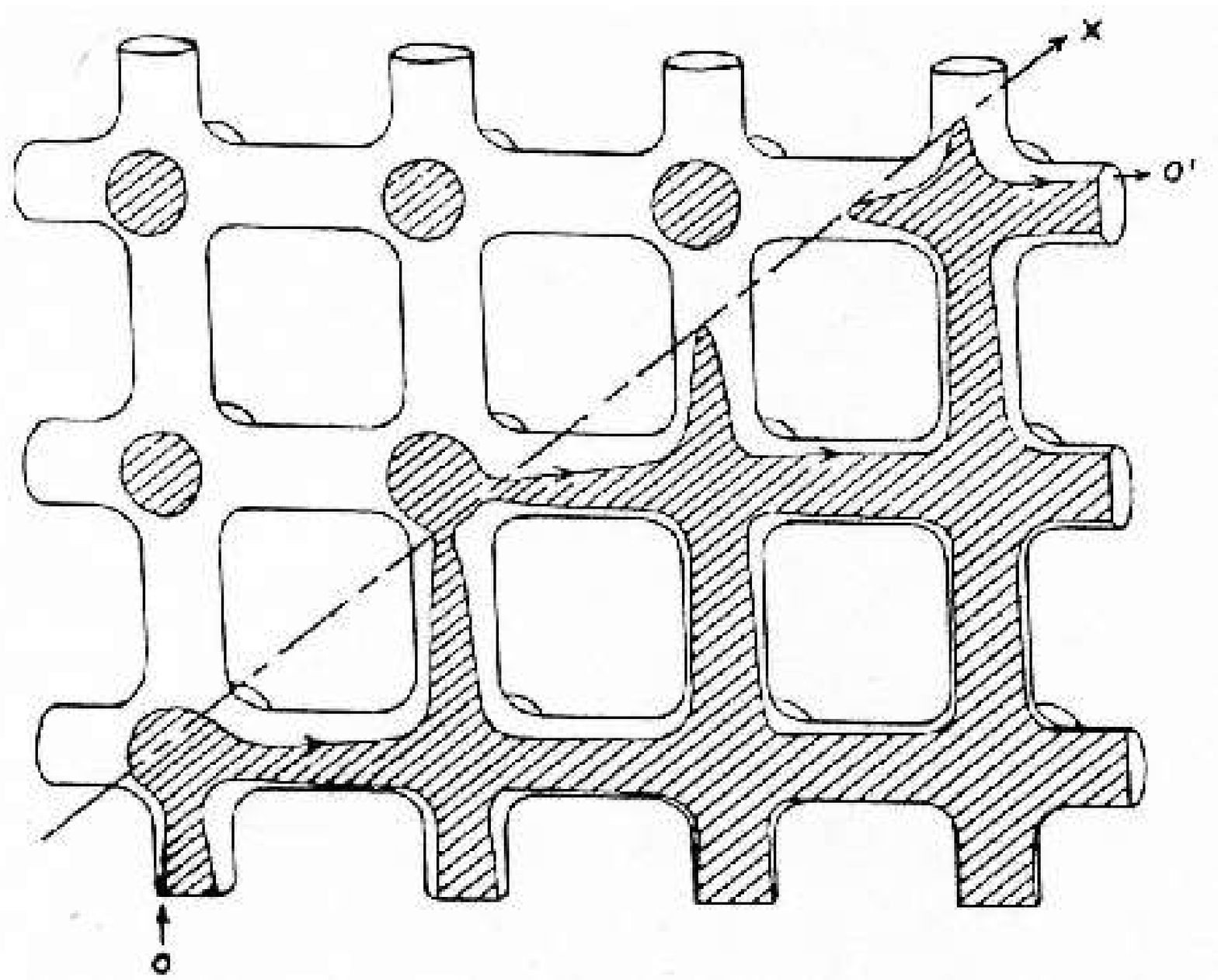}
\cr
(a)&&(b)\cr
\epsfxsize=3.75cm\epsfbox{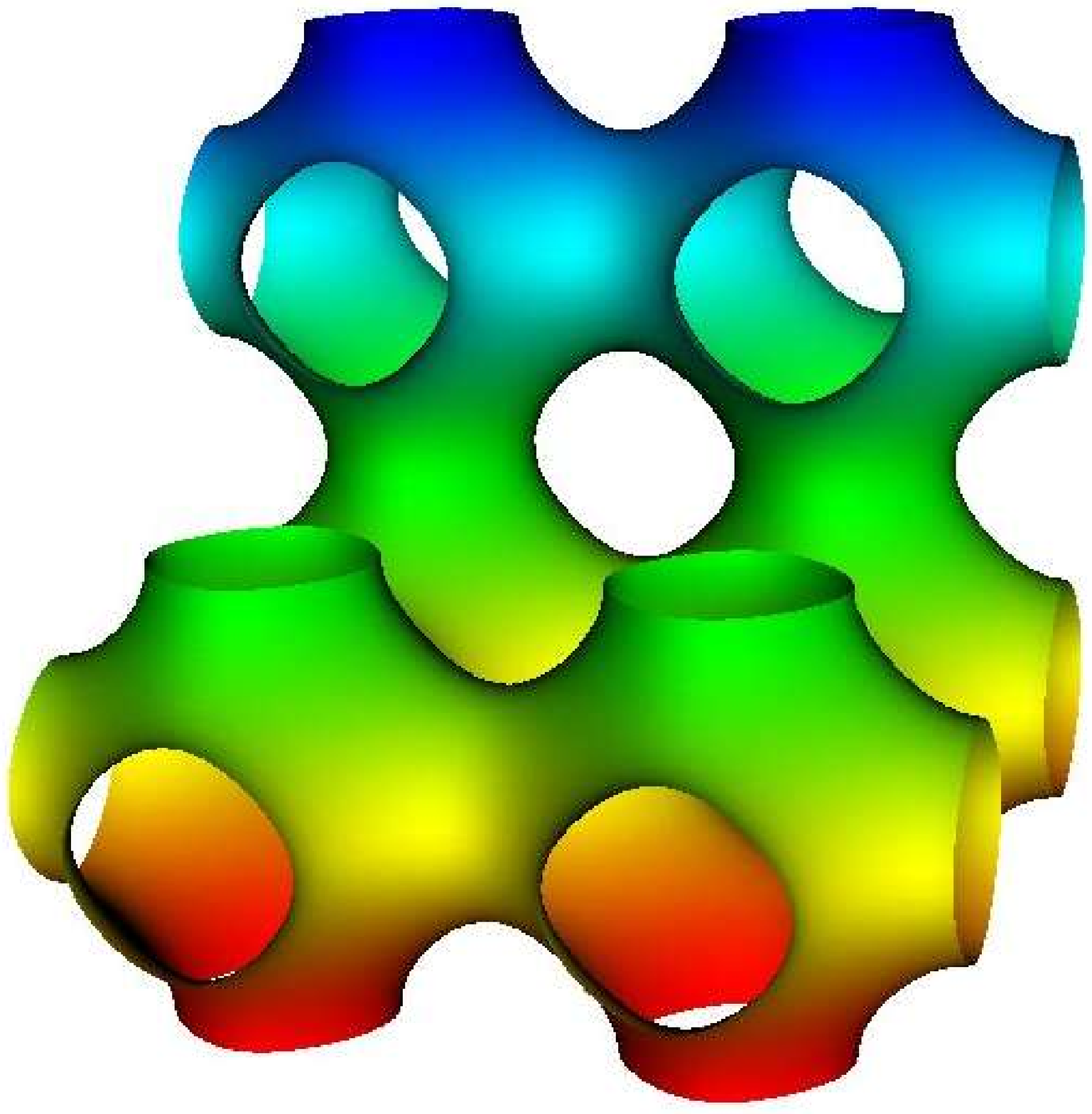}
&&
\epsfxsize=3.75cm\epsfbox{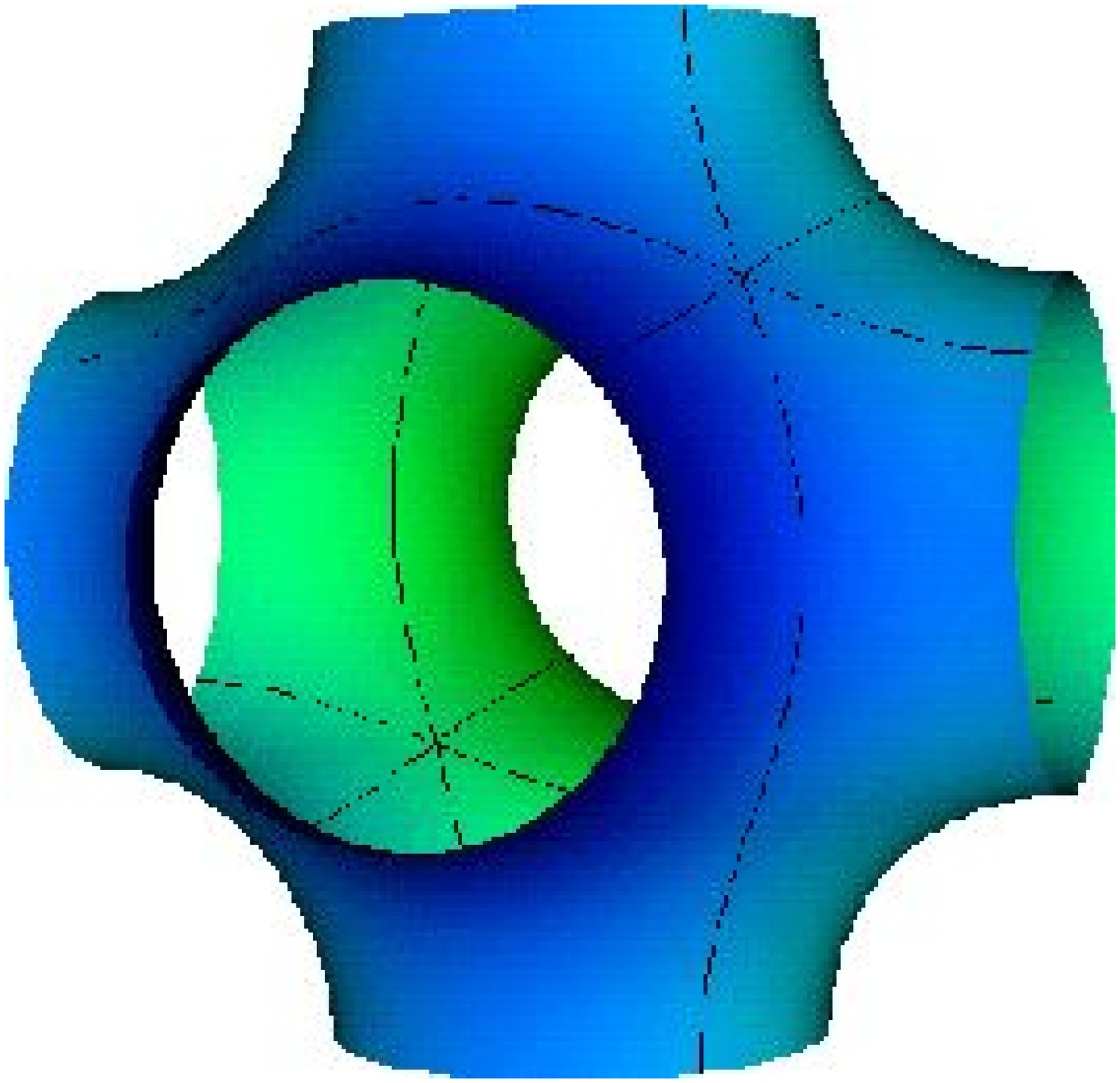}
\cr
(c)&&(d)\cr
}}$$
\caption{%
(a) Rough sketch of the stability zones for a genus$-3$ surface \cite{LAK56}; 
(b) Example of section of a genus$-3$ surface; 
(c) Component of the ``prison bars'' surface; 
(d) Basic component of the ``prison bars'' surface with basic cycles shown on it 
(each repeated twice).}
\label{fig:genus3}
\end{figure}
Our software core is a C++ library called NTC (Novikov Torus Conjecture) built over 
the graphics library VTK \cite{SML98} that provides the basic calls to build isosurfaces 
meshes and perform elementary geometric and topological operations.
What our library adds to VTK is mainly the capability of dealing with ``periodic geometry''
and critical slices (e.g. planes that cut a surfaces in a saddle point) and the capability
of evaluating topological quantities like homology classes of loops.
\par
The library has been released under the GPL license and it is downladable at the 
address http://ntc.sf.net/.
\section{A toy model}
\label{sec:g3}
It is well known \cite{AM76} that a first rough expression for topologically non-trivial
FS can be quickly obtained through the {\sl tight-binding} approximation.
\par
For a simple cubic crystal this approximation leads in the lowest order to the following FF 
\cite{Dav84}:
$$\cE(\bp)=\cos(2\pi p_x)+\cos(2\pi p_y)+\cos(2\pi p_z)\,.$$
It turns out that this FF is the simplest non-trivial case for our system:
indeed its level surfaces are either spheres, when $E<-1$ or $E>1$, or 
genus-3 rank-3 surfaces shaped as a a sort of ``three-dimensional prison bars'' 
(fig.~\ref{fig:genus3}).
\par
\begin{figure}[tl]
\epsfxsize=9cm\epsfbox[20 60 575 575]{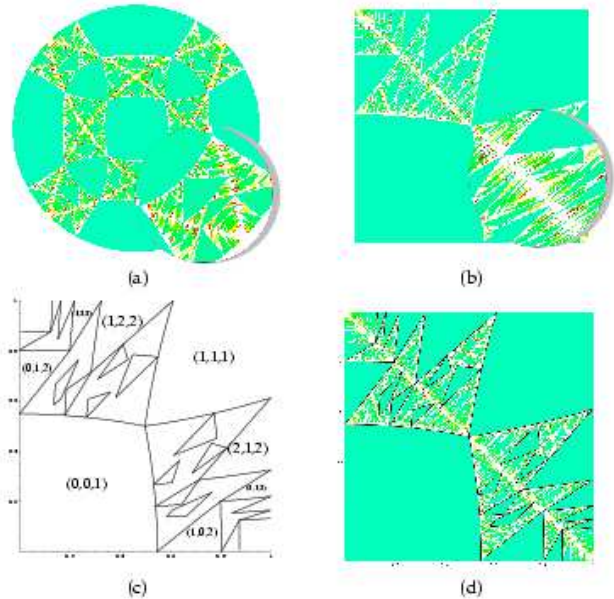}
\caption{%
(a) SM for $\cE=\cos(2\pi p_x)+\cos(2\pi p_y)+\cos(2\pi p_z)$; 
(b) detail of the fractal in coordinates $(H_x,H_y,1)$; 
(c) boundaries and Miller indexes of a few stability zones evaluated analytically; 
(d) comparison between numerical and analytical calculations.}
\label{fig:g3frac}
\end{figure}
Several efforts have been made in Sixties to understand the topology of orbits in
this elementary case, but the production of SM did not get more detail than the rough 
sketch shown in fig.\ref{fig:genus3}(a). 
\par
This function is a perfect ``toy-model'' for testing our algorithm
since its analytical expression is so simple to make possible verifying many things 
analytically. Moreover, it satisfies the following property: its level surfaces 
$\{\cE=c\}$ and $\{\cE=-c\}$ differ only by a translation, so that a magnetic field
$\bH$ that gives rise to open orbits at energy $c$ does the same at the opposite energy, 
i.e. all energy intervals giving rise to open orbits have the form $[-e(\bH),e(\bH)]$;
open orbits therefore arise for {\sl every} direction of $\bH$ at the level $e=0$,
namely the SM at the zero level is identical to the ``global SM''. It is easy to
verify (even by hand!) that this SM has more than a zone and therefore (sec.\ref{sec:top}) 
it has infinitely many zones distributed in a fractal-like way.
Finally, at the zero-energy level it is possible to get a simple analytical expression 
for the saddle points as function of the magnetic field direction, an information that 
is enough to obtain, in principle, the analytical expression of the boundary for
any island.
\par
\begin{figure}
\begin{tabular}{|c|c|c|c|}
\hline
\multicolumn{4}{|c|}{Biggest zones in the SM for}\\
\multicolumn{4}{|c|}{$\cE=\cos(p_x)+\cos(p_y)+\cos(p_z)$}\\
\hline
Miller index&Area&Miller index&Area\\
\hline
$(0,0,1)$&$(2.83\pm.02)10^{-1}$&$(1,5,5)$&$(4.1\pm.2)10^{-3}$\\
$(1,1,1)$&$(2.03\pm.01)10^{-1}$&$(2,5,8)$&$(4.1\pm.4)10^{-3}$\\
$(1,2,2)$&$(8.2\pm.2)10^{-2}$&$(2,6,7)$&$(3.4\pm.4)10^{-3}$\\
$(0,1,2)$&$(5.1\pm.1)10^{-2}$&$(4,7,8)$&$(3.0\pm.3)10^{-3}$\\
$(1,3,3)$&$(2.1\pm.1)10^{-2}$&$(0,3,4)$&$(2.9\pm.4)10^{-3}$\\
$(2,3,4)$&$(1.7\pm.1)10^{-2}$&$(3,5,7)$&$(2.7\pm.3)10^{-3}$\\
$(1,3,5)$&$(9.6\pm.5)10^{-3}$&$(1,6,6)$&$(2.0\pm.1)10^{-3}$\\
$(1,4,6)$&$(9.6\pm.5)10^{-3}$&$(4,5,8)$&$(2.0\pm.4)10^{-3}$\\
$(0,2,3)$&$(9.0\pm.6)10^{-3}$&$(5,8,10)$&$(1.9\pm.4)10^{-3}$\\
$(2,4,5)$&$(8.6\pm.6)10^{-3}$&$(4,6,9)$&$(1.8\pm.3)10^{-3}$\\
$(1,4,4)$&$(8.3\pm.3)10^{-3}$&$(1,6,10)$&$(1.7\pm.1)10^{-3}$\\
$(1,2,4)$&$(6.2\pm.5)10^{-3}$&$(5,9,11)$&$(1.6\pm.2)10^{-3}$\\
$(3,4,6)$&$(4.7\pm.5)10^{-3}$&$(4,6,7)$&$(1.5\pm.2)10^{-3}$\\
\hline
\end{tabular}
\caption{Miller indices associated to the biggest zones of the SM in the unitary
square (fig.\ref{fig:g3frac}).}
\label{fig:tab}
\end{figure}
We produced the SM in the square $0\leq H_{x,y}\leq1$, $H_z=1$ by evaluating the Miller
index associated to every direction in a $10^3\times10^3$ equally spaced square lattice 
(fig.\ref{fig:g3frac}(b)) and then obtained the whole SM by symmetry (fig.\ref{fig:g3frac}(a)).
In fig.\ref{fig:g3frac}(c) are shown the boundaries of the biggest zones obtained
directly from the analytical expression of the saddle points; as it is possible to verify
from fig.\ref{fig:g3frac}(d), there is a perfect agreement between numerical and analytical
data. Similar calculations made for a piecewise smooth quadratic function with the same 
symmetries of $\cE$ showed a similar agreement even at energies different from zero.
\par
Finally, in fig.\ref{fig:tab} we list the Miller indices corresponding to the biggest 26 zones 
together with an estimate of their area in the unitary square. A thorough description of 
the study of this FS can be found in [DL03]\cite{DL03}. 
\section{Numerical exploration for Gold and Silver}
\begin{figure}
$$\vbox{\halign{\hfill#\hfill&#\hskip .1cm&\hfill#\hfill\cr
\epsfxsize=4.6cm\epsfbox[20 100 575 575]{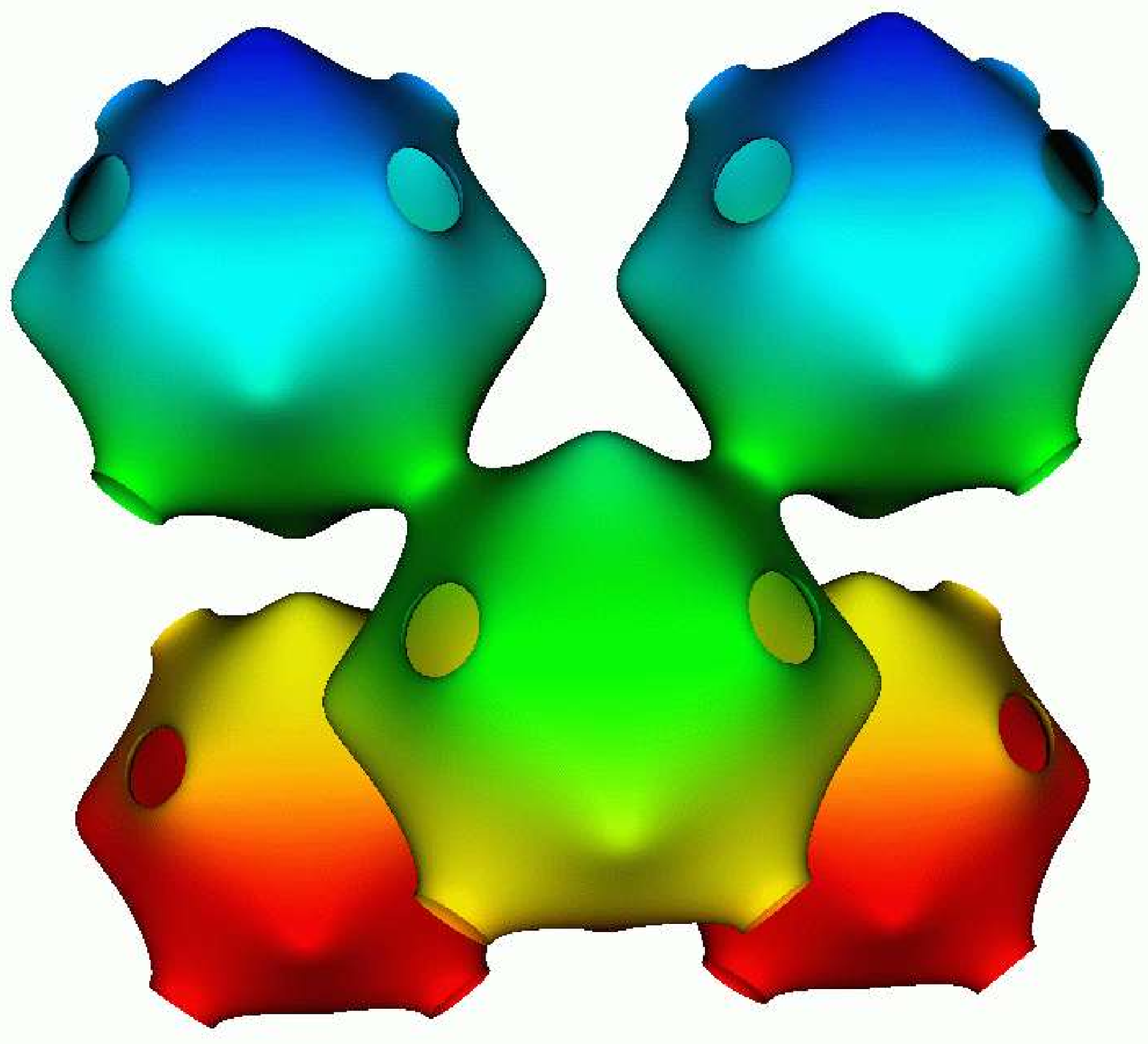}
&&
\scalebox{.75}{
\pspicture(0,0)(5,5)
\rput[lb](0,0){\epsfxsize=5.cm\epsfbox{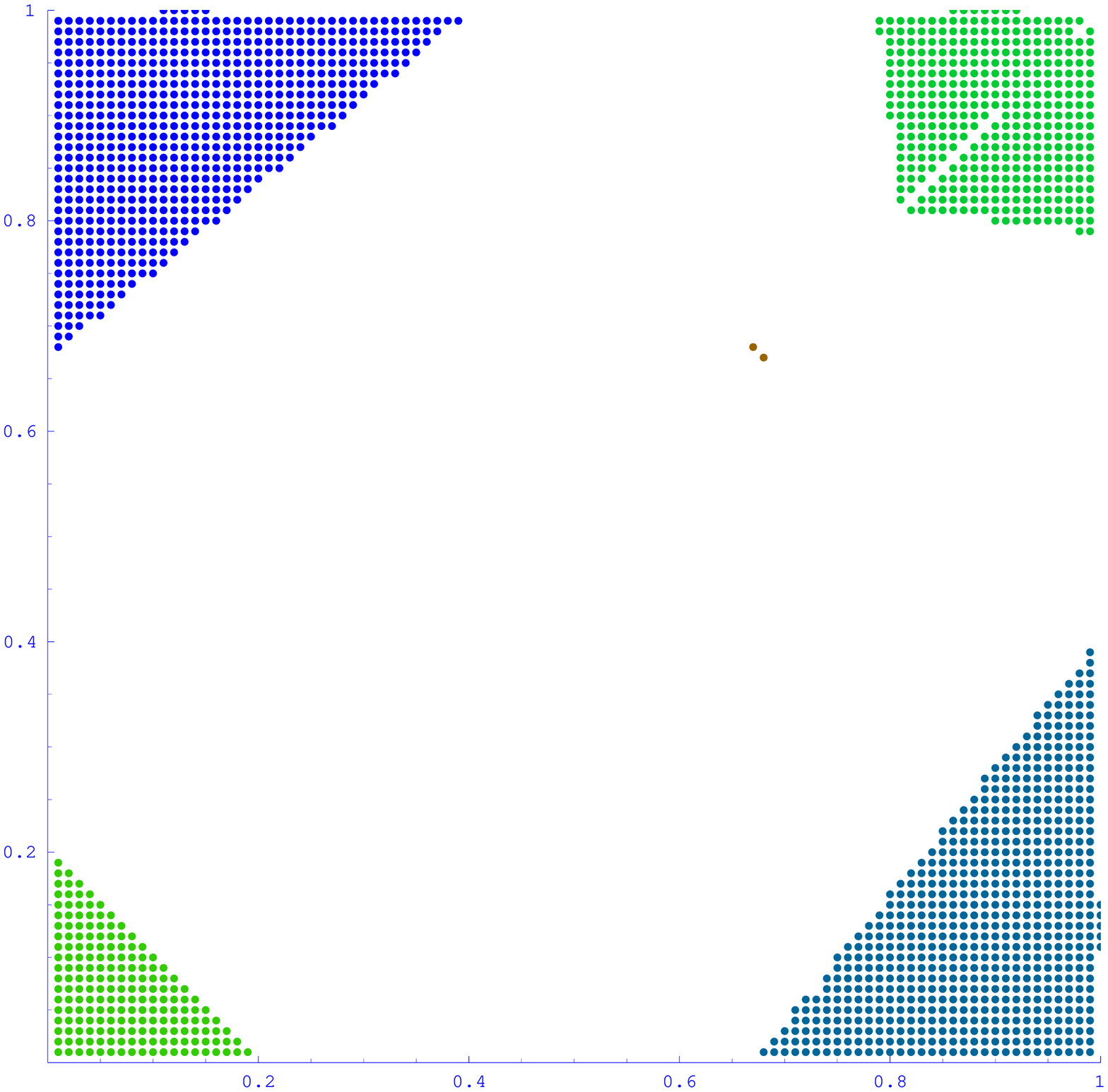}}
\rput[lb](.75,.75){\scalebox{.7}{$(0,0,1)$}}
\rput[lb](1,3.5){\scalebox{.7}{$(0,1,1)$}}
\rput[lb](3,1){\scalebox{.7}{$(1,0,1)$}}
\rput[lb](3,4.3){\scalebox{.7}{$(1,1,1)$}}
\rput[c](3.5,3){\scalebox{.7}{$(3,3,4)$}}
\endpspicture}
\cr
\noalign{\vskip .4cm}	
(a)&&(b)\cr
\noalign{\vskip .4cm}	
\epsfxsize=4.25cm\epsfbox[310 130 560 300]{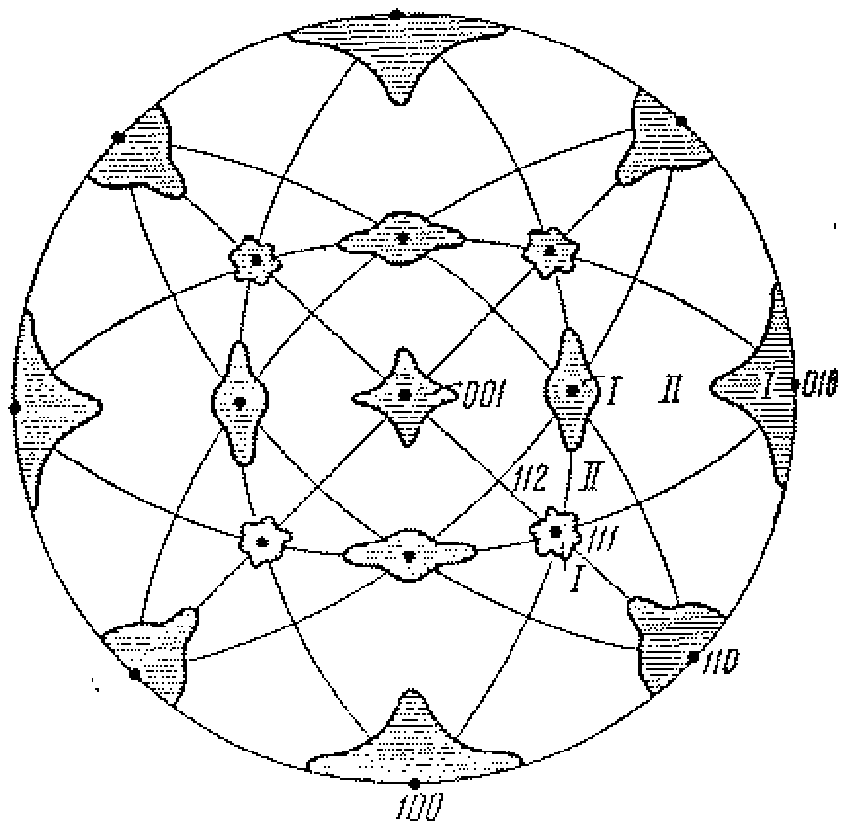}
&&
\epsfxsize=3.7cm\epsfbox{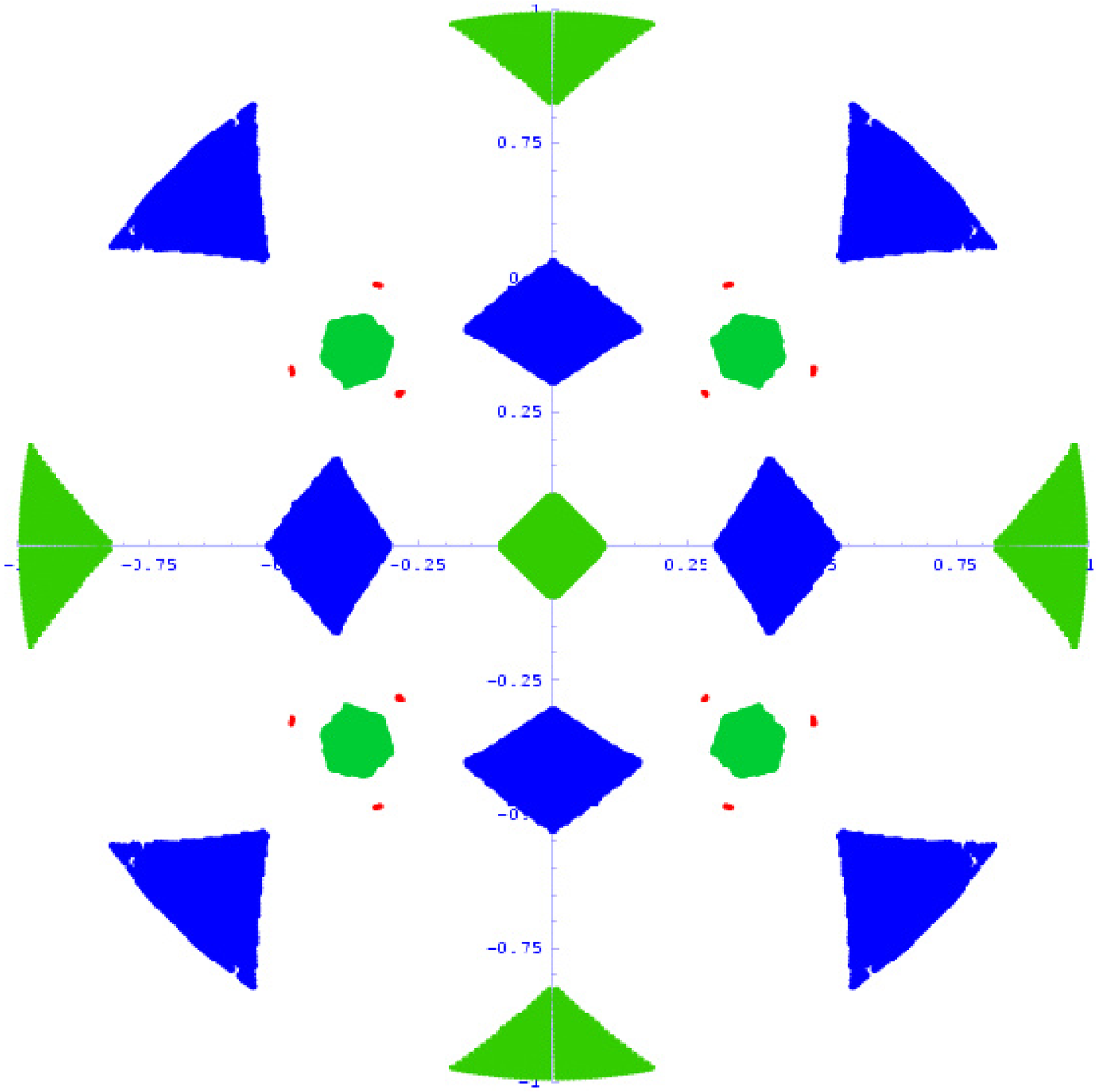}\cr
\noalign{\vskip .4cm}	
(c)&&(d)\cr
}}$$
\caption{%
(a) FS of Au according to the Halse formula; 
(b) numerical SM of Au in the unit square in coordinates $(H_x,H_y,1)$
obtained using the NTC library; (c) experimental SM for Au obtained experimentally by 
Gaidukov \cite{Ga60}; (d) numerical SM for Au obtained from (b) by symmetry.}
\label{fig:Au}
\end{figure}
\begin{figure}
$$\vbox{\halign{\hfill#\hfill&#\hskip .15cm&\hfill#\hfill\cr
\epsfxsize=4.6cm\epsfbox[20 100 575 575]{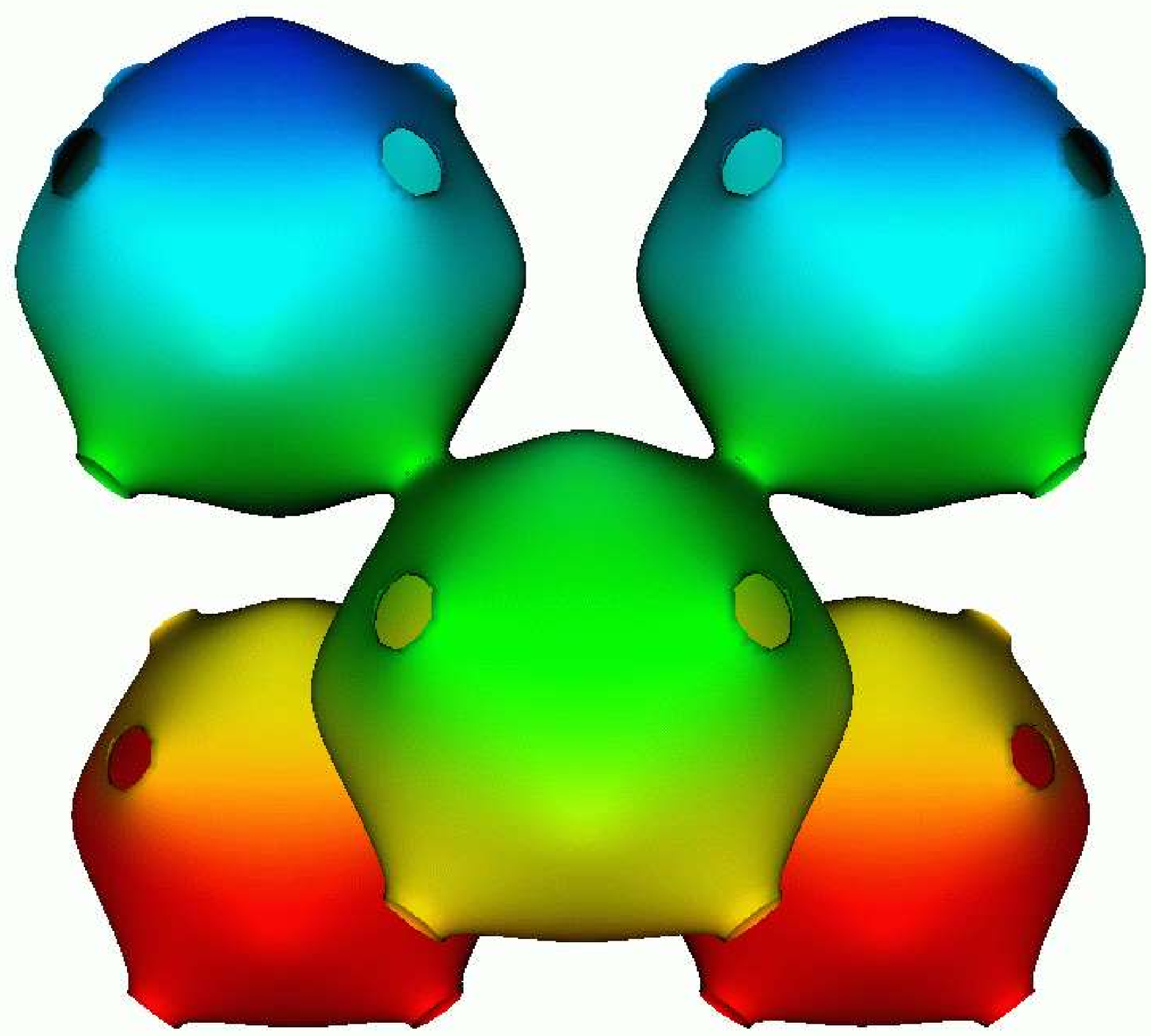}
&&
\scalebox{.75}{
\pspicture(0,0)(5,5)
\rput[lb](0,0){\epsfxsize=5.cm\epsfbox{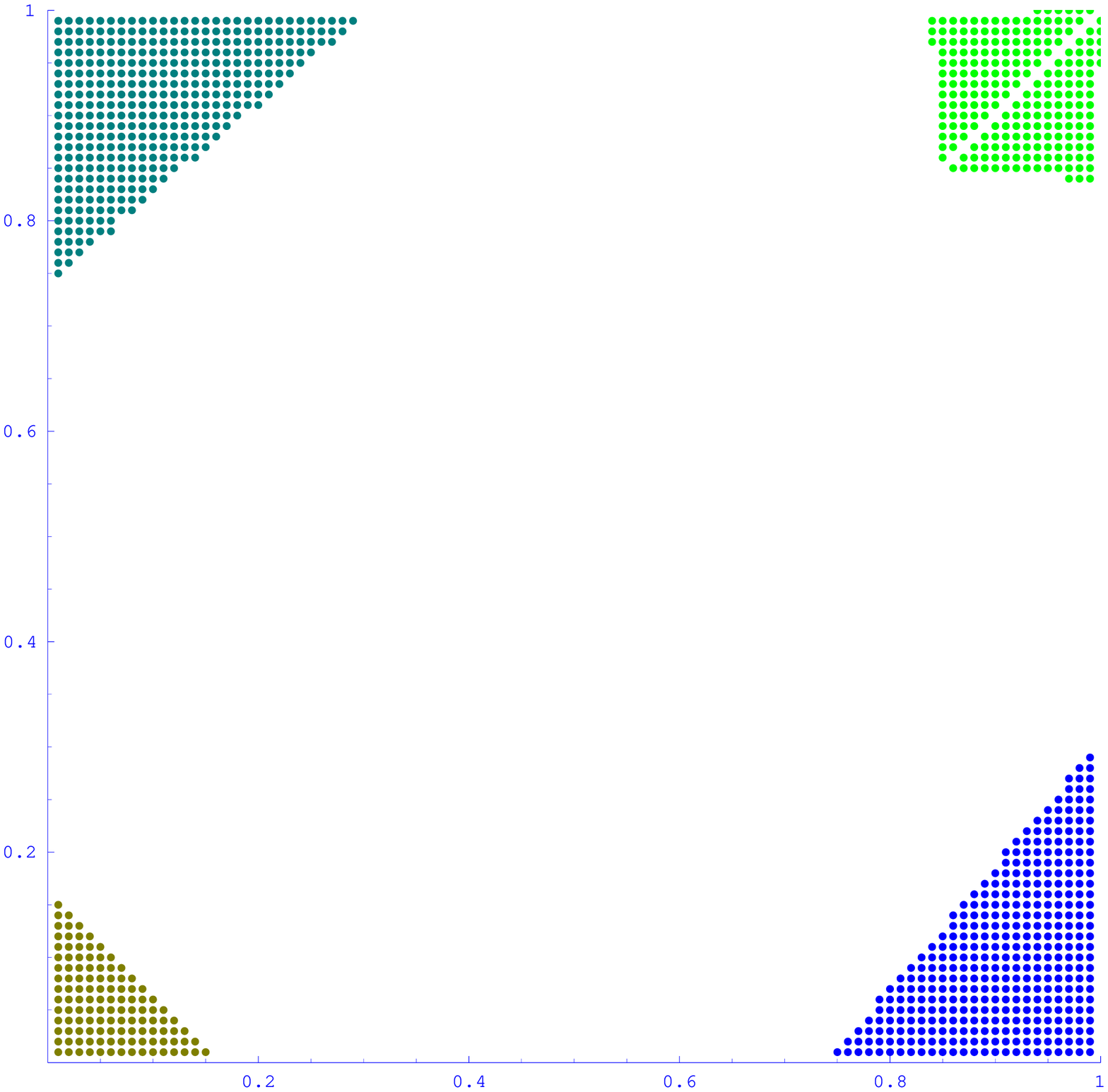}}
\rput[lb](.65,.65){\scalebox{.7}{$(0,0,1)$}}
\rput[lb](.7,3.8){\scalebox{.7}{$(0,1,1)$}}
\rput[lb](3.2,.8){\scalebox{.7}{$(1,0,1)$}}
\rput[lb](3.2,4.3){\scalebox{.7}{$(1,1,1)$}}
\endpspicture
}
\cr
\noalign{\vskip .4cm}	
(a)&&(b)\cr
\noalign{\vskip .4cm}	
\epsfxsize=4.6cm\epsfbox[-20 45 575 558]{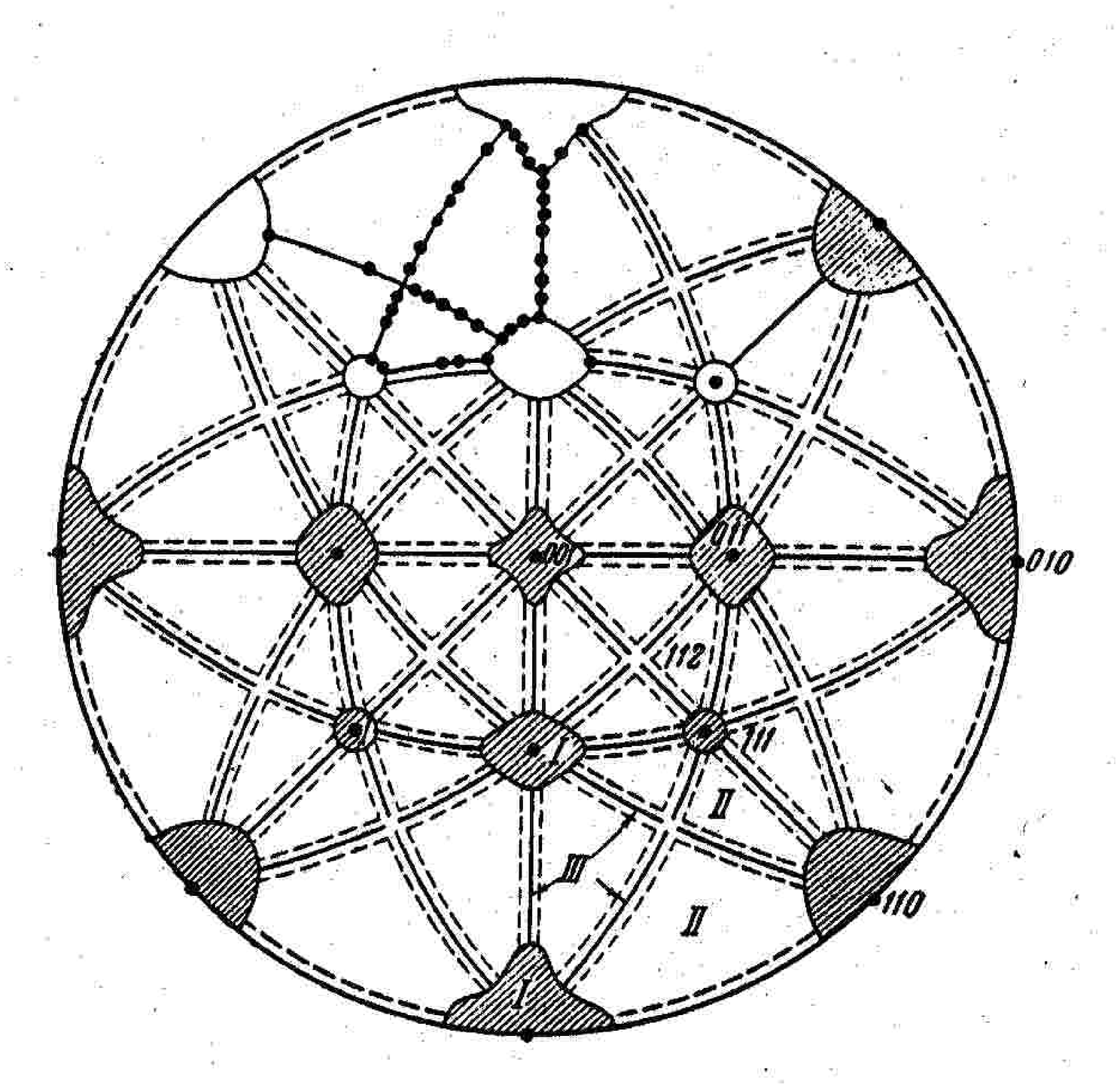}
&&
\epsfxsize=3.7cm\epsfbox{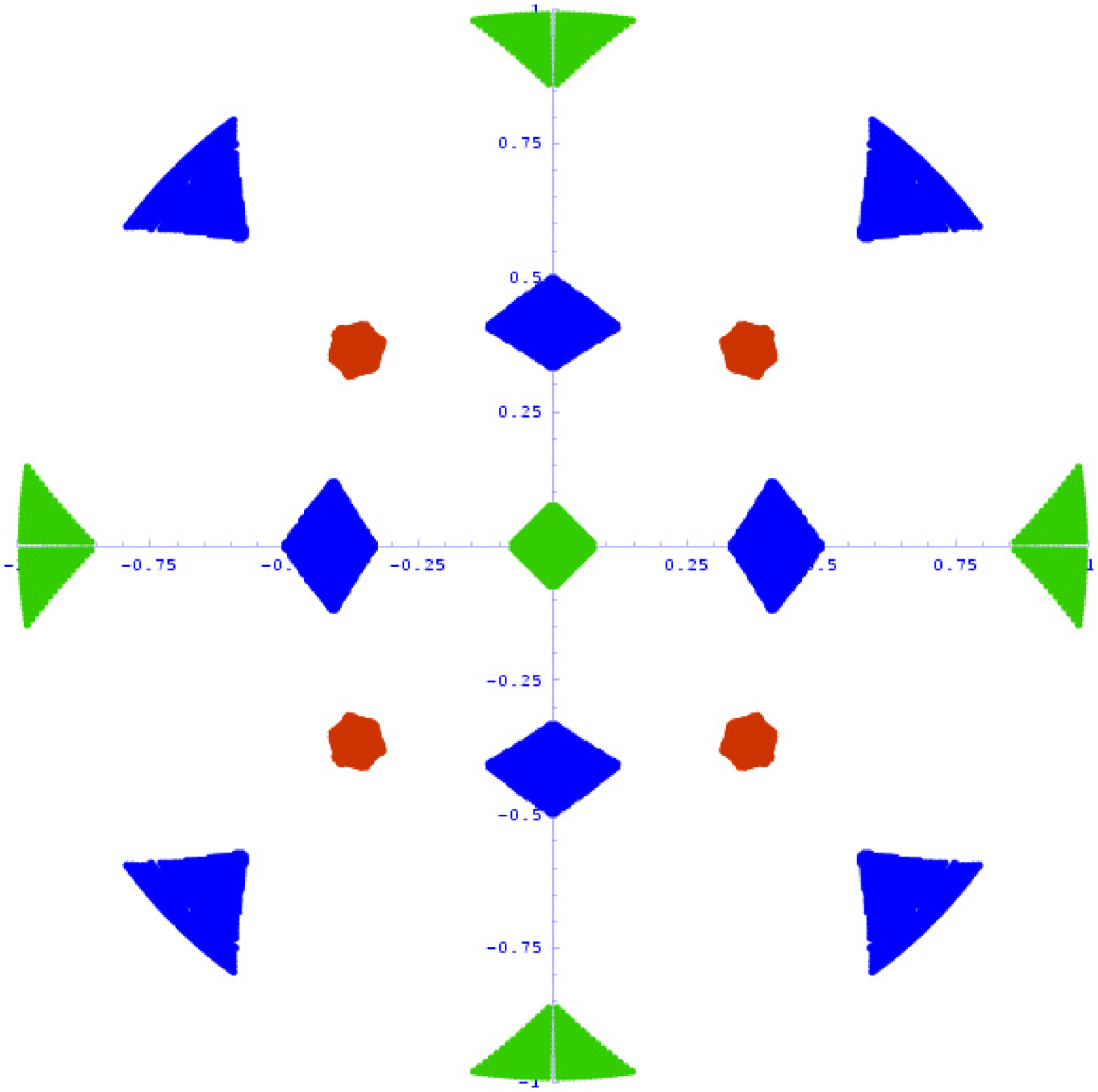}\cr
\noalign{\vskip .4cm}	
(c)&&(d)\cr
}}$$
\caption{%
(a) FS of Ag according to the Halse formula; 
(b) numerical SM of Ag in the unit square in coordinates $(H_x,H_y,1)$
obtained using the NTC library; (c) experimental SM for Ag obtained experimentally by 
Gaidukov and Alekseevskii\cite{AG62c}; (d) numerical SM for Ag obtained from (b) by symmetry.}
\label{fig:Ag}
\end{figure}
%
It is a well-established result that Gold and Silver have a genus-4 FS: both FS are spheres 
with four handles directed along the four diagonals of the cube (e.g. see fig.\ref{fig:Au}(a)).
\par
Extremely precise estimates of the FF values can now be achieved through numerical
calculations, for example using advanced tight-binding methods \cite{PM03}, but
several facts lead us to use older estimates in form of analytical approximations:
on one hand, Gaidukov experiments were conducted with a magnetic field intensity of 
$H\simeq1T$, barely around the minimum intensity for the FS topology to be relevant,
so that a priori we cannot expect more than a rough agreement with the experimental
data; on the other hand, dealing with FF known only numerically would make so  
complicated and slow our algorithm that it would make sense doing it only once 
it is clear that the algorithm works properly. Since the numerical results proved to 
be very encouraging, we are currently working on adding to our algorithm the possibility to
work on FF provided in numerical form and it is our hope that, now that numerical 
algorithms exist to compare theoretical predictions and experimental results, new 
experiments will be made with stronger magnetic fields to allow detailed comparisons.
\par
Very precise analytical approximations for the FF of Au, Ag and Cu have been known 
since late Fifties; we extracted the ones we used from Halse's formulas \cite{Hal69} 
that are known to give rise to surfaces within $.2\%$ from the actual Fermi Surface:
$$\vbox{\halign{$#$\hfil&$#$\hfil\cr
	\cE(p) &= 3 - \sum\cos(2\pi p_x)\cos(2\pi p_y)\cr
		       &+\alpha\left(3-\sum\cos\left(4\pi p_x\right)\right)\cr
		       &+\beta\left(3-\sum\cos\left(4\pi p_x\right)\cos\left(2\pi p_y\right)\cos\left(2\pi p_z\right)\right)\cr
		       &+\gamma\left(3-\sum\cos\left(4\pi p_x\right)\cos\left(4\pi p_y\right)\right)\cr
\noalign{\noindent$+\delta\left(6-\sum\left(\cos\left(6\pi p_x\right)\cos\left(2\pi p_y\right)-\cos\left(2\pi p_x\right)\cos\left(6\pi p_y\right)\right)\right)$}
}}
$$
$$
\vbox{
  \tabskip=5pt
  \halign{#\hfill&\hfill$#$\hfill&\hfill$#$\hfill&\hfill$#$\hfill&\hfill$#$\hfill&\hfill$#$\hfill\cr
	    &\alpha&\beta&\gamma&\delta&E_F\cr
	Cu&\phantom{-}.00693&-.42501&-.01679&-.03772&\phantom{-}1.69167\cr
	Ag&-.12030&-.90187&-.14086&-.09483&-0.89789\cr
	Au&-.16635&-1.25516&-.09914&-.12704&-2.26213\cr
}}
$$
Since FS have genus four, in the first Brillouin zone there can be at most three cylinders of
closed orbits and therefore no more than a pair of ``warped planes'', like in the genus-3 case,
but this time there are six (rather than four) basic cycles.
Exactly like in the ``prison bars'' case, the symmetries of the FS are such to allow to 
restrict the numerical exploration to half of the square $[0,1]\times[0,1]$
representing the directions $\bH=(H_x,H_y,1)$.
We performed numerical simulations for Gold and Silver sampling the unit square
with a lattice of $100\times100$ points. The resolution we used for generating 
a simplicial decomposition of the Fermi Surface was again a $100\times100$
lattice in the cube. The symmetry of the picture with respect to the diagonal of the 
first quadrant is a useful extra check about the correctness of the output we get.
\par
The biggest four zones have been found in both Gold and Silver and are
labeled exactly by the direction at their centers, namely $(0,0,1)$, $(1,0,1)$, $(0,1,1)$ 
and $(1,1,1)$ (as it had to be according to Dynnikov\cite{Dyn99}). 
\par
A fifth zone labeled by $(3,3,4)$ has been found in Gold only; it does not contain the 
direction with its label and is the only one that has not been detected in Gaidukov's 
experiment.
Since this zone is very small it is not clear whether it has not been seen because
of the limits of the semiclassical approximation or of the low intensity of the
magnetic field used in Sixties by Gaidukov. New experimental data taken with a 
stronger magnetic field should be able to answer this question.
In fig.\ref{fig:Au}(d) and \ref{fig:Ag}(d) we extend by symmetry those data and show the 
resulting stereographic projection comparing it with the ones generated by Gaidukov from 
its experimental data.
In both cases the pictures are in very good agreement.
%
\section{Conclusions}
The method described in this paper allows, for the first time, to
predict the Stereographic Map of a metal from the knowledge of its Fermi Surface.
\par
Applying this method we first verified, with a toy model, Dynnikov's discovery of the 
fractal structures of ``global SM'' and then produced SM for Au and Ag based on analytical 
expressions for their FS available in literature. The zones found numerically for the 
toy model agree perfectly with the corresponding boundaries found analytically; 
the ones found for Au and Ag are very close to the corresponding experimental data,
that is even more remarkable considering that those data, obtained more than forty 
years ago, were taken just at the lower threshold for the semiclassical 
approximation to hold.
\par
A new zone, not detected experimentally, was found in Au. It is not clear whether
this is only due to unaccuracies in the FS utlized in calculations or rather to 
the too weak magnetic field used in the Gaidukov experiment or possibly to purely
quantum mechanical corrections to the semiclassical approximation.
It is our hope new accurate Stereographic Maps for Au will be produced to help 
finding out the right answer.
\section{Acknowledgments}
We wish to thank S.P. Novikov for introducing the subject and precious scientific
discussions. We also thank I. Dynnikov for suggesting the method we implemented and for
stimulating discussions that were essential for the present work, 
I. Mazin for his scientific help and advice and P. Ruggerone for critically reading this work.
All numerical calculations were made with computers kindly provided by INFN 
(www.ca.infn.it) and IPST (www.ipst.umd.edu).
We also acknowledge financial support from Indam (indam.mat.uniroma1.it).
\medskip
\bibliography{genus4} 

\begin{thebibliography}{40}
\expandafter\ifx\csname natexlab\endcsname\relax\def\natexlab#1{#1}\fi
\expandafter\ifx\csname bibnamefont\endcsname\relax
  \def\bibnamefont#1{#1}\fi
\expandafter\ifx\csname bibfnamefont\endcsname\relax
  \def\bibfnamefont#1{#1}\fi
\expandafter\ifx\csname citenamefont\endcsname\relax
  \def\citenamefont#1{#1}\fi
\expandafter\ifx\csname url\endcsname\relax
  \def\url#1{\texttt{#1}}\fi
\expandafter\ifx\csname urlprefix\endcsname\relax\def\urlprefix{URL }\fi
\providecommand{\bibinfo}[2]{#2}
\providecommand{\eprint}[2][]{\url{#2}}

\bibitem[{\citenamefont{Justi and Scheffers}(1937)}]{JS37}
\bibinfo{author}{\bibfnamefont{E.}~\bibnamefont{Justi}} \bibnamefont{and}
  \bibinfo{author}{\bibfnamefont{H.}~\bibnamefont{Scheffers}},
  \bibinfo{journal}{Phys. Z.} \textbf{\bibinfo{volume}{38}},
  \bibinfo{pages}{891} (\bibinfo{year}{1937}).

\bibitem[{\citenamefont{Lifschitz et~al.}(1956)\citenamefont{Lifschitz, Azbel,
  and Kaganov}}]{LAK56}
\bibinfo{author}{\bibfnamefont{I.}~\bibnamefont{Lifschitz}},
  \bibinfo{author}{\bibfnamefont{M.}~\bibnamefont{Azbel}}, \bibnamefont{and}
  \bibinfo{author}{\bibfnamefont{M.}~\bibnamefont{Kaganov}},
  \bibinfo{journal}{JETP} \textbf{\bibinfo{volume}{3}}, \bibinfo{pages}{143}
  (\bibinfo{year}{1956}).

\bibitem[{\citenamefont{Gaidukov}(1960)}]{Ga60}
\bibinfo{author}{\bibfnamefont{Y.}~\bibnamefont{Gaidukov}},
  \bibinfo{journal}{JETP} \textbf{\bibinfo{volume}{10}}, \bibinfo{pages}{913}
  (\bibinfo{year}{1960}).

\bibitem[{\citenamefont{Alexeevskii and Gaidukov}(1962{\natexlab{a}})}]{AG62c}
\bibinfo{author}{\bibfnamefont{N.}~\bibnamefont{Alexeevskii}} \bibnamefont{and}
  \bibinfo{author}{\bibfnamefont{Y.}~\bibnamefont{Gaidukov}},
  \bibinfo{journal}{JETP} \textbf{\bibinfo{volume}{15}}, \bibinfo{pages}{49}
  (\bibinfo{year}{1962}{\natexlab{a}}).

\bibitem[{\citenamefont{Alexeevskii and Gaidukov}(1959)}]{AG59}
\bibinfo{author}{\bibfnamefont{N.}~\bibnamefont{Alexeevskii}} \bibnamefont{and}
  \bibinfo{author}{\bibfnamefont{Y.}~\bibnamefont{Gaidukov}},
  \bibinfo{journal}{JETP} \textbf{\bibinfo{volume}{9}}, \bibinfo{pages}{311}
  (\bibinfo{year}{1959}).

\bibitem[{\citenamefont{Alexeevskii et~al.}(1961)\citenamefont{Alexeevskii,
  Gaidukov, Lifschitz, and Peschanskii}}]{AGLP61}
\bibinfo{author}{\bibfnamefont{N.}~\bibnamefont{Alexeevskii}},
  \bibinfo{author}{\bibfnamefont{Y.}~\bibnamefont{Gaidukov}},
  \bibinfo{author}{\bibfnamefont{I.}~\bibnamefont{Lifschitz}},
  \bibnamefont{and}
  \bibinfo{author}{\bibfnamefont{V.}~\bibnamefont{Peschanskii}},
  \bibinfo{journal}{JETP} \textbf{\bibinfo{volume}{12}}, \bibinfo{pages}{837}
  (\bibinfo{year}{1961}).

\bibitem[{\citenamefont{Alexeevskii and Gaidukov}(1960)}]{AG60}
\bibinfo{author}{\bibfnamefont{N.}~\bibnamefont{Alexeevskii}} \bibnamefont{and}
  \bibinfo{author}{\bibfnamefont{Y.}~\bibnamefont{Gaidukov}},
  \bibinfo{journal}{JETP} \textbf{\bibinfo{volume}{10}}, \bibinfo{pages}{481}
  (\bibinfo{year}{1960}).

\bibitem[{\citenamefont{Alexeevskii and Gaidukov}(1962{\natexlab{b}})}]{AG62a}
\bibinfo{author}{\bibfnamefont{N.}~\bibnamefont{Alexeevskii}} \bibnamefont{and}
  \bibinfo{author}{\bibfnamefont{Y.}~\bibnamefont{Gaidukov}},
  \bibinfo{journal}{JETP} \textbf{\bibinfo{volume}{14}}, \bibinfo{pages}{256}
  (\bibinfo{year}{1962}{\natexlab{b}}).

\bibitem[{\citenamefont{Alexeevskii and Gaidukov}(1962{\natexlab{c}})}]{AG62b}
\bibinfo{author}{\bibfnamefont{N.}~\bibnamefont{Alexeevskii}} \bibnamefont{and}
  \bibinfo{author}{\bibfnamefont{Y.}~\bibnamefont{Gaidukov}},
  \bibinfo{journal}{JETP} \textbf{\bibinfo{volume}{14}}, \bibinfo{pages}{770}
  (\bibinfo{year}{1962}{\natexlab{c}}).

\bibitem[{\citenamefont{Alexeevskii and Gaidukov}(1963)}]{AG63}
\bibinfo{author}{\bibfnamefont{N.}~\bibnamefont{Alexeevskii}} \bibnamefont{and}
  \bibinfo{author}{\bibfnamefont{Y.}~\bibnamefont{Gaidukov}},
  \bibinfo{journal}{JETP} \textbf{\bibinfo{volume}{16}}, \bibinfo{pages}{1481}
  (\bibinfo{year}{1963}).

\bibitem[{\citenamefont{Alexeevskii et~al.}(1964)\citenamefont{Alexeevskii,
  Karstens, and Mozhaev}}]{AKM64}
\bibinfo{author}{\bibfnamefont{N.}~\bibnamefont{Alexeevskii}},
  \bibinfo{author}{\bibfnamefont{E.}~\bibnamefont{Karstens}}, \bibnamefont{and}
  \bibinfo{author}{\bibfnamefont{V.}~\bibnamefont{Mozhaev}},
  \bibinfo{journal}{JETP} \textbf{\bibinfo{volume}{19}}, \bibinfo{pages}{1979}
  (\bibinfo{year}{1964}).

\bibitem[{\citenamefont{Pippard}(1957)}]{Pip57}
\bibinfo{author}{\bibfnamefont{A.}~\bibnamefont{Pippard}},
  \bibinfo{journal}{Phil. Trans. Roy. Soc. A} \textbf{\bibinfo{volume}{250}},
  \bibinfo{pages}{325} (\bibinfo{year}{1957}).

\bibitem[{\citenamefont{Lifschitz et~al.}(1957)\citenamefont{Lifschitz, Azbel,
  and Kaganov}}]{LAK57}
\bibinfo{author}{\bibfnamefont{I.}~\bibnamefont{Lifschitz}},
  \bibinfo{author}{\bibfnamefont{M.}~\bibnamefont{Azbel}}, \bibnamefont{and}
  \bibinfo{author}{\bibfnamefont{M.}~\bibnamefont{Kaganov}},
  \bibinfo{journal}{JETP} \textbf{\bibinfo{volume}{4}}, \bibinfo{pages}{41}
  (\bibinfo{year}{1957}).

\bibitem[{\citenamefont{Lifschitz et~al.}(1973)\citenamefont{Lifschitz, Azbel,
  and Kaganov}}]{LAK73}
\bibinfo{author}{\bibfnamefont{I.}~\bibnamefont{Lifschitz}},
  \bibinfo{author}{\bibfnamefont{M.}~\bibnamefont{Azbel}}, \bibnamefont{and}
  \bibinfo{author}{\bibfnamefont{M.}~\bibnamefont{Kaganov}},
  \emph{\bibinfo{title}{Electron theory of metals}}
  (\bibinfo{publisher}{Consultants Bureau}, \bibinfo{year}{1973}).

\bibitem[{\citenamefont{Lifschitz and Peschanskii}(1959)}]{LP59}
\bibinfo{author}{\bibfnamefont{I.}~\bibnamefont{Lifschitz}} \bibnamefont{and}
  \bibinfo{author}{\bibfnamefont{V.}~\bibnamefont{Peschanskii}},
  \bibinfo{journal}{JETP} \textbf{\bibinfo{volume}{8}}, \bibinfo{pages}{875}
  (\bibinfo{year}{1959}).

\bibitem[{\citenamefont{Lifschitz and Peschanskii}(1960)}]{LP60}
\bibinfo{author}{\bibfnamefont{I.}~\bibnamefont{Lifschitz}} \bibnamefont{and}
  \bibinfo{author}{\bibfnamefont{V.}~\bibnamefont{Peschanskii}},
  \bibinfo{journal}{JETP} \textbf{\bibinfo{volume}{11}}, \bibinfo{pages}{137}
  (\bibinfo{year}{1960}).

\bibitem[{\citenamefont{Chambers}(1960)}]{Cha60}
\bibinfo{author}{\bibfnamefont{R.}~\bibnamefont{Chambers}}, in
  \emph{\bibinfo{booktitle}{Magnetoresistance}}, edited by
  \bibinfo{editor}{\bibnamefont{Harrison}} \bibnamefont{and}
  \bibinfo{editor}{\bibnamefont{Webb}} (\bibinfo{year}{1960}).

\bibitem[{\citenamefont{Chambers}(1957)}]{Cha57}
\bibinfo{author}{\bibfnamefont{R.}~\bibnamefont{Chambers}},
  \bibinfo{journal}{Proc. Roy. Soc.} \textbf{\bibinfo{volume}{238}},
  \bibinfo{pages}{344} (\bibinfo{year}{1957}).

\bibitem[{\citenamefont{Shoenberg}(1960)}]{Sho60}
\bibinfo{author}{\bibfnamefont{D.}~\bibnamefont{Shoenberg}},
  \bibinfo{journal}{Proc. Mag.} \textbf{\bibinfo{volume}{5}},
  \bibinfo{pages}{105} (\bibinfo{year}{1960}).

\bibitem[{\citenamefont{Shoenberg}(1962)}]{Sho62}
\bibinfo{author}{\bibfnamefont{D.}~\bibnamefont{Shoenberg}},
  \bibinfo{journal}{Proc. Trans. Roy. Soc. A} \textbf{\bibinfo{volume}{255}},
  \bibinfo{pages}{85} (\bibinfo{year}{1962}).

\bibitem[{\citenamefont{Pippard}(1989)}]{Pip89}
\bibinfo{author}{\bibfnamefont{A.}~\bibnamefont{Pippard}},
  \emph{\bibinfo{title}{Magnetoresistance in Metals}}
  (\bibinfo{publisher}{CUP}, \bibinfo{year}{1989}).

\bibitem[{\citenamefont{Novikov}(1982)}]{Nov82}
\bibinfo{author}{\bibfnamefont{S.}~\bibnamefont{Novikov}},
  \bibinfo{journal}{Usp. Mat. Nauk (RMS)} \textbf{\bibinfo{volume}{37:5}},
  \bibinfo{pages}{3} (\bibinfo{year}{1982}).

\bibitem[{\citenamefont{Novikov and Maltsev}(1998)}]{NM98}
\bibinfo{author}{\bibfnamefont{S.}~\bibnamefont{Novikov}} \bibnamefont{and}
  \bibinfo{author}{\bibfnamefont{A.}~\bibnamefont{Maltsev}},
  \bibinfo{journal}{Usp. Fiz. Nauk} \textbf{\bibinfo{volume}{41:3}},
  \bibinfo{pages}{231} (\bibinfo{year}{1998}).

\bibitem[{\citenamefont{Zorich}(1984)}]{Zor84}
\bibinfo{author}{\bibfnamefont{A.}~\bibnamefont{Zorich}},
  \bibinfo{journal}{Usp. Mat. Nauk (RMS)} \textbf{\bibinfo{volume}{39:5}},
  \bibinfo{pages}{235} (\bibinfo{year}{1984}).

\bibitem[{\citenamefont{Dynnikov}(1992)}]{Dyn92}
\bibinfo{author}{\bibfnamefont{I.}~\bibnamefont{Dynnikov}},
  \bibinfo{journal}{Usp. Mat. Nauk (RMS)} \textbf{\bibinfo{volume}{57:3}},
  \bibinfo{pages}{172} (\bibinfo{year}{1992}).

\bibitem[{\citenamefont{Dynnikov}(1997)}]{Dyn97}
\bibinfo{author}{\bibfnamefont{I.}~\bibnamefont{Dynnikov}},
  \bibinfo{journal}{AMS Transl} \textbf{\bibinfo{volume}{179}},
  \bibinfo{pages}{45} (\bibinfo{year}{1997}).

\bibitem[{\citenamefont{Dynnikov}(1999)}]{Dyn99}
\bibinfo{author}{\bibfnamefont{I.}~\bibnamefont{Dynnikov}},
  \bibinfo{journal}{RMS} \textbf{\bibinfo{volume}{54:1}}, \bibinfo{pages}{21}
  (\bibinfo{year}{1999}).

\bibitem[{\citenamefont{Kapitza}(1929)}]{Kap29}
\bibinfo{author}{\bibfnamefont{P.}~\bibnamefont{Kapitza}},
  \bibinfo{journal}{Proc. Roy. Soc.} \textbf{\bibinfo{volume}{125A}},
  \bibinfo{pages}{292} (\bibinfo{year}{1929}).

\bibitem[{\citenamefont{Peierls}(1931)}]{Pei31}
\bibinfo{author}{\bibfnamefont{R.}~\bibnamefont{Peierls}},
  \bibinfo{journal}{Ann. Phys.} \textbf{\bibinfo{volume}{10}},
  \bibinfo{pages}{193} (\bibinfo{year}{1931}).

\bibitem[{\citenamefont{Ashcroft and Mermin}(1976)}]{AM76}
\bibinfo{author}{\bibfnamefont{N.}~\bibnamefont{Ashcroft}} \bibnamefont{and}
  \bibinfo{author}{\bibfnamefont{N.}~\bibnamefont{Mermin}},
  \emph{\bibinfo{title}{Solid State Physics}} (\bibinfo{publisher}{North
  Holland}, \bibinfo{year}{1976}).

\bibitem[{\citenamefont{Novikov and Maltsev}(2003)}]{NM03}
\bibinfo{author}{\bibfnamefont{S.}~\bibnamefont{Novikov}} \bibnamefont{and}
  \bibinfo{author}{\bibfnamefont{A.}~\bibnamefont{Maltsev}},
  \bibinfo{journal}{J. of Statistical Physics} \textbf{\bibinfo{volume}{115}},
  \bibinfo{pages}{31} (\bibinfo{year}{2003}), \bibinfo{note}{cond-mat/0312708}.

\bibitem[{\citenamefont{Lifschitz and Kaganov}(1980)}]{LK80}
\bibinfo{author}{\bibfnamefont{I.}~\bibnamefont{Lifschitz}} \bibnamefont{and}
  \bibinfo{author}{\bibfnamefont{A.}~\bibnamefont{Kaganov}}, in
  \emph{\bibinfo{booktitle}{The Fermi Surface}}, edited by
  \bibinfo{editor}{\bibfnamefont{M.}~\bibnamefont{SpringFord}}
  (\bibinfo{publisher}{CUP}, \bibinfo{year}{1980}).

\bibitem[{\citenamefont{Abrikosov}(1988)}]{Abr88}
\bibinfo{author}{\bibfnamefont{A.}~\bibnamefont{Abrikosov}},
  \emph{\bibinfo{title}{Fundamentals of the Theory of Metals}}
  (\bibinfo{publisher}{North Holland}, \bibinfo{year}{1988}).

\bibitem[{\citenamefont{Lifschitz and Kaganov}(1960)}]{LK60}
\bibinfo{author}{\bibfnamefont{I.}~\bibnamefont{Lifschitz}} \bibnamefont{and}
  \bibinfo{author}{\bibfnamefont{A.}~\bibnamefont{Kaganov}},
  \bibinfo{journal}{Soviet Physics} \textbf{\bibinfo{volume}{2}},
  \bibinfo{pages}{831} (\bibinfo{year}{1960}).

\bibitem[{\citenamefont{Novikov et~al.}(1989)\citenamefont{Novikov, Dubrovin,
  and Fomenko}}]{NDF89}
\bibinfo{author}{\bibfnamefont{S.}~\bibnamefont{Novikov}},
  \bibinfo{author}{\bibfnamefont{B.}~\bibnamefont{Dubrovin}}, \bibnamefont{and}
  \bibinfo{author}{\bibfnamefont{A.}~\bibnamefont{Fomenko}},
  \emph{\bibinfo{title}{Modern Geometry III}} (\bibinfo{publisher}{Springer
  Verlag}, \bibinfo{year}{1989}).

\bibitem[{\citenamefont{Schroeder et~al.}(1998)\citenamefont{Schroeder, Martin,
  and Lorense}}]{SML98}
\bibinfo{author}{\bibfnamefont{W.}~\bibnamefont{Schroeder}},
  \bibinfo{author}{\bibfnamefont{K.}~\bibnamefont{Martin}}, \bibnamefont{and}
  \bibinfo{author}{\bibfnamefont{B.}~\bibnamefont{Lorense}},
  \emph{\bibinfo{title}{The Visualization Toolkit}}
  (\bibinfo{publisher}{Prentice Hall PTR}, \bibinfo{year}{1998}).

\bibitem[{\citenamefont{Davydov}(1984)}]{Dav84}
\bibinfo{author}{\bibfnamefont{A.}~\bibnamefont{Davydov}},
  \emph{\bibinfo{title}{Th\'eorie du solide}} (\bibinfo{publisher}{MIR},
  \bibinfo{year}{1984}).

\bibitem[{\citenamefont{Leo}(2003)}]{DL03}
\bibinfo{author}{\bibfnamefont{R.~D.} \bibnamefont{Leo}},
  \bibinfo{journal}{SIADS} \textbf{\bibinfo{volume}{2:4}}, \bibinfo{pages}{517}
  (\bibinfo{year}{2003}),
  \urlprefix\url{http://epubs.siam.org/sam-bin/dbq/article/40664}.

\bibitem[{\citenamefont{Papaconstantopoulos and Mehl}(2003)}]{PM03}
\bibinfo{author}{\bibfnamefont{D.}~\bibnamefont{Papaconstantopoulos}}
  \bibnamefont{and} \bibinfo{author}{\bibfnamefont{M.}~\bibnamefont{Mehl}},
  \bibinfo{journal}{Journal of Physics: Condensed Matter}
  \textbf{\bibinfo{volume}{15}}, \bibinfo{pages}{413} (\bibinfo{year}{2003}).

\bibitem[{\citenamefont{Halse}(1969)}]{Hal69}
\bibinfo{author}{\bibfnamefont{M.}~\bibnamefont{Halse}},
  \bibinfo{journal}{Phil. Trans. Roy. Soc. London A}
  \textbf{\bibinfo{volume}{265}}, \bibinfo{pages}{507} (\bibinfo{year}{1969}).

\end{thebibliography}
\end{document}